\documentclass[aps,prl,reprint,twocolumn,floatfix,superscriptaddress,longbibliography]{revtex4-2}

\pdfoutput=1

\usepackage{amsfonts,amsmath,amssymb}

\usepackage{graphicx}
\graphicspath{{graphics/}} 

\usepackage{xcolor}
\usepackage[colorlinks,citecolor=blue,linkcolor=blue,urlcolor=blue]{hyperref}

\usepackage{lmodern} 
\usepackage[T1]{fontenc}
\usepackage[utf8]{inputenc} 
\usepackage{nicefrac}
\usepackage{textcomp}
\usepackage{braket}
\usepackage{slashed}
\usepackage{tensor}
\usepackage[subnum]{cases}
\usepackage{mathtools}
\usepackage{bm}

\usepackage{standalone}

\usepackage{amsthm}
\usepackage{amsfonts, amsmath, amssymb}
\usepackage{graphicx}
\usepackage{dcolumn}
\usepackage{bm}
\usepackage{color}
\usepackage[utf8]{inputenc}
\usepackage[colorlinks,citecolor=blue,linkcolor=blue,urlcolor=blue]{hyperref}
\usepackage{tikz}
\usepackage{braket}
\usepackage{physics}
\usepackage{color}
\usepackage{mathtools}
\usepackage{float}
\usepackage{subfigure}
\usepackage{stmaryrd} 
\usepackage{enumerate}

\makeatletter
\def\maketitle{
	\@author@finish
	\title@column\titleblock@produce
	\suppressfloats[t]}
\makeatother

\newcommand{\h}{\hbar}

\newcommand{\w}{\omega}
\newcommand{\W}{\Omega}

\newcommand{\tld}[1]{\widetilde{#1}}
\newcommand{\hsp}{\hspace{.5cm}}

\begin{document}

\title{Exact Duality at Low Energy in a Josephson Tunnel Junction\\
Coupled to a Transmission Line}

\author{Luca Giacomelli}
\affiliation{Université Paris Cité, CNRS, Matériaux et Phénomènes Quantiques, 75013 Paris, France}
\author{Michel H. Devoret}
\affiliation{Physics Department, University of California Santa Barbara, California 93106, USA}
\affiliation{Google Quantum AI, 301 Mentor Dr, Goleta, California 93111, USA}
\author{Cristiano Ciuti}
\affiliation{Université Paris Cité, CNRS, Matériaux et Phénomènes Quantiques, 75013 Paris, France}

\begin{abstract}
We theoretically explore the low-energy behavior of a Josephson tunnel junction coupled to a finite-length, charge-biased transmission line and compare it to its flux-biased counterpart. For transmission lines of increasing length, we show that the low-energy charge-dependent energy bands of the charge-biased configuration can be exactly mapped onto those of the flux-biased system via a well-defined duality transformation of circuit parameters. In the limit of an infinitely long transmission line, the influence of boundary conditions vanishes, and both circuits reduce to a resistively shunted Josephson junction. This convergence reveals the system’s intrinsic self-duality and critical behavior. Our exact formulation of charge-flux duality provides a foundation for generalizations to more complex superconductor-insulator phase transitions.
\end{abstract}

\maketitle

{\it Introduction ---}
In a pioneering work, Schmid \cite{schmid1983diffusion} predicted a localization-delocalization transition for a quantum Brownian particle in a periodic cosine potential. It was also proposed \cite{schmid1983diffusion,schon1990quantum} that this open quantum system could be physically realized using a superconducting Josephson tunnel junction coupled to a resistive environment. Specifically, when the resistance exceeds the quantum of resistance, $R_q = h/(2e)^2$, the superconducting phase becomes localized. Remarkably, the transition point was predicted to be independent from the ratio $E_J/E_C$ between Josephson inductive energy and capacitive charging energy \cite{bulgadaev1984phase}. This surprising prediction was based on an \textit{approximate charge-flux duality}, which would map the small-junction regime ($E_J \ll E_C$) into the large-junction regime ($E_J \gg E_C$). In other words, the regime whose physics is dominated by Cooper pair charge tunneling can be mapped into the regime of flux-tunneling (also known as phase slip) \cite{guinea1985diffusion}. In the original context of the quantum Brownian particle, the mapping relates the two regimes of a weakly bound and a tightly bound particle in a periodic potential \cite{weiss2012quantum}.

With the tremendous progress in Josephson junction technology, circuit quantum electrodynamics (QED) \cite{blais2021circuit}, and the development of high-impedance quantum circuits and transmission-line resonators \cite{sundaresan2015beyond,forn2016ultrastrong,bosman2017multimode,puertas2019tunable,kuzmin2019superstrong,leger2019observation,kuzmin2021inelastic,mehta2023downconversion,crescini2023evidence,leger2023revealing,fraudet2025direct}, the Schmid transition has recently become the focus of intense debate, driven by a series of experimental \cite{murani2020absence,subero2023bolometric,kuzmin2025observation} and theoretical \cite{hakonen2021comment,murani2021reply,houzet2020critical,masuki2022absence,sepulcre2022comment,masuki2022reply,daviet2023nature,houzet2024microwave,kashuba2024limitations,paris2024resilience,kurilovich2025quantum} studies, some of which doubt the existence of the Schmid transition and the validity of the approximate duality. While the existence of a critical point was unambiguously confirmed \cite{giacomelli2024emergent,paris2024resilience}, a fundamental open question is wether a duality relationship exists, also for intermediate values of the ratio $E_J/E_C$ and when the resistive environment is replaced by a finite-size transmission line. Establishing an exact duality would provide a powerful framework for understanding the emergence of critical points also in more complex quantum circuits and superconducting–insulating phase transitions~\cite{fazio2001quantum}.

\begin{figure*}[t!]
  \centering
  \includegraphics[width=0.9 \textwidth]{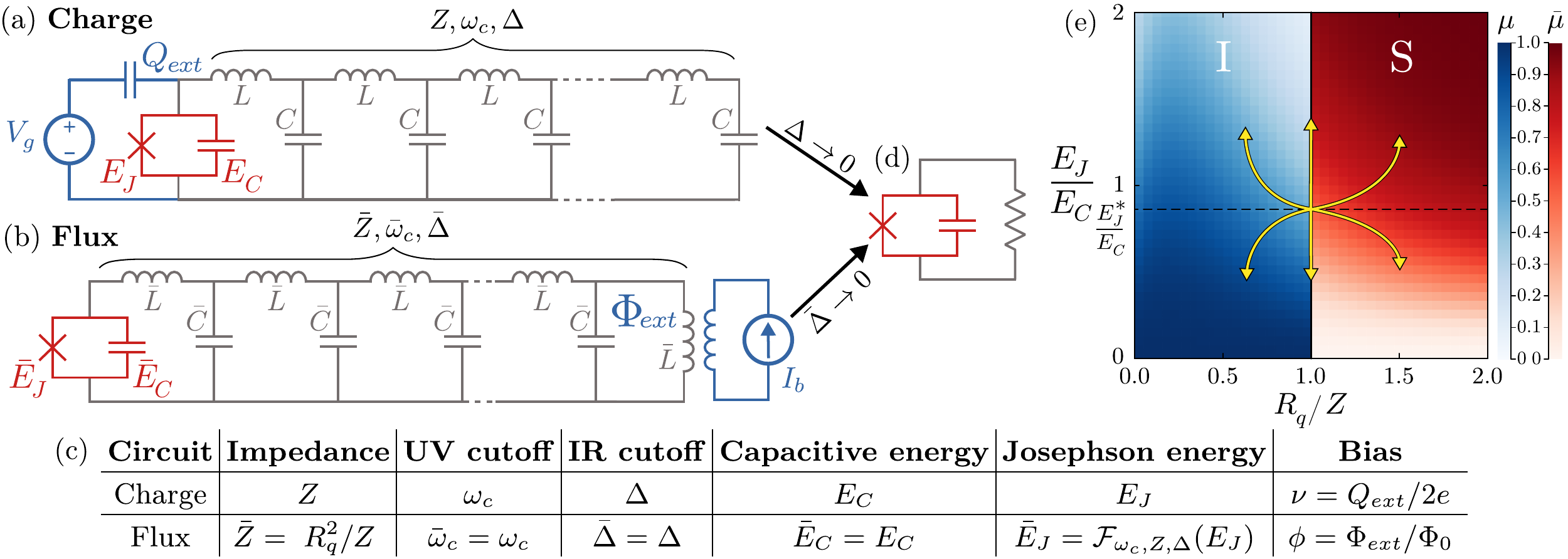}
  \caption{(a) \textit{Charge} circuit: a Josephson tunnel junction with capacitive energy $E_C=e^2/2C_J$ and Josephson energy $E_J$, coupled to a finite-length transmission line characterized by inductances $L$, capacitances $C$ and number of modes $N_m$ or, equivalently, by impedance $Z=\sqrt{L/C}$, cutoff frequency $\omega_c=2/\sqrt{LC}$, and low energy mode spacing $\Delta=\pi/(N_m\sqrt{LC})$. The circuit includes a superconducting island and can be biased via gate charge $\nu = Q_{ext}/2e = V_g C_g / 2e$.
(b) \textit{Flux} circuit: a shorted transmission line without a superconducting island, forming a superconducting loop, that can biased by an external magnetic flux $\phi = \Phi_{ext}/\Phi_0$, where $\Phi_0$ is the flux quantum for a Cooper pair.
(c) Table summarizing the parameter correspondences between the two circuits, such that that their low-energy spectra—specifically, energy band dispersions vs. normalized bias—are identical under the dual mapping. The transformation $\mathcal{F}_{\omega_c,Z,\Delta}$ of $E_J$ is discussed in the main text.
(d) In the infinite-length limit, boundary conditions become irrelevant and both circuits reduce to a resistively shunted Josephson junction.
(e) Schmid phase diagram showing insulating (I) and superconducting (S) phases. The two heatmaps show parameters $\mu$ and $\bar\mu$ that characterize the spectra of the two circuits (defined later in the text) for $\hbar\omega_c=4E_C$ and $\hbar\Delta=E_C$. Points at dual impedances with the same value of these parameters have the same low-energy spectrum, and are hence dual. Arrows indicate this duality transformation for three couples of points.}
  \label{fig:circuits}
\end{figure*}

In this letter, we demonstrate an exact charge–flux duality that holds across the entire range of $E_J/E_C$. Using exact diagonalization, we solve two finite-length realizations of the Ohmic Caldeira–Leggett model: a capacitively terminated and an inductively terminated transmission line (Fig.~\ref{fig:circuits}(a)–(b)). In contrast to most studies that probe the Schmid transition by perturbatively evaluating a response function in the limiting cases of infinite transmission-line length, infinite ultraviolet cutoff, zero temperature, and zero frequency, we circumvent the subtleties associated with these idealized limits \cite{schmid1983diffusion,bulgadaev1984phase,guinea1985diffusion,schon1990quantum}.
 Instead, we characterize the phase transition by exactly computing the low-energy spectrum of finite-length systems as a function of a bias parameter. As the line length increases, the charge-dependent low-energy bands of the first circuit map exactly onto the flux-dependent bands of the second through a transformation that exchanges low and high impedances and rescales $E_J$, while leaving all other parameters unchanged (Fig.~\ref{fig:circuits}(c),(e)). The two circuits are therefore rigorously dual at low energies. In the infinite-length limit, boundary conditions become irrelevant and the two circuits become indistinguishable (Fig.~\ref{fig:circuits}(d)), rendering the system \textit{self-dual} \cite{weiss2012quantum}. 

Our results extend the conventional notion of approximate self-duality into an exact mapping valid across the entire parameter space. In particular, both circuits exhibit  scale invariance at the impedance $Z=R_q$, and the full Schmid critical line maps onto itself (Fig.~\ref{fig:circuits}(e)). This demonstrates conclusively that the phase transition is independent of the ratio $E_J/E_C$, a feature difficult to capture with other methods \cite{daviet2023nature}, and circumvents recently raised concerns regarding the extendedness or compactness of the phase variable -- which would imply the absence of a transition \cite{murani2020absence} -- as well as the potential influence of dangerously irrelevant terms in renormalization-group analyses, which could alter the phase diagram \cite{masuki2022absence}.

{\it Breakdown of Classical Circuit Duality with Josephson Elements ---}  In classical circuit theory, the dual of a circuit is one that satisfies the same equations with voltages and currents interchanged. One finds that inductances and capacitances are swapped, as well as parallel and series configurations. Specifically, an open-circuited (capacitively terminated) lumped-element transmission line (gray part of Fig.~\ref{fig:circuits}(a)) is dual to a short-circuited (inductively terminated) transmission line (gray part of Fig.~\ref{fig:circuits}(b)).
This duality extends to the quantum regime, where charge fluctuations in one circuit of characteristic impedance $Z=\sqrt{L/C}$ correspond to phase fluctuations in its dual counterpart of impedance $\bar{Z} = \sqrt{\bar L/\bar C}=R_q^2/Z$. In the limit of an infinitely long transmission line, these two configurations can be interpreted as distinct realizations of the Caldeira-Leggett model, describing an admittance and an impedance, respectively, with a key distinction in their impedance at zero frequency \cite{vool2017introduction,supp}. However, when a Josephson tunnel junction is coupled to these transmission lines, as depicted in Fig.~\ref{fig:circuits}, the conventional circuit duality is broken. This is because the dual circuit would necessarily contain the dual of a Josephson tunnel junction, where charge tunneling is replaced by phase tunneling, i.e. a phase-slip element \cite{supp,mooij2006superconducting}.  

Remarkably, as we now demonstrate, an exact numerical solution reveals an exact low-energy duality between these two circuits, despite the apparent breakdown of the conventional duality framework in the presence of a Josephson element. In other words, at low energies, the Josephson tunnel junction and the phase-slip junction in the circuits of Fig.~\ref{fig:circuits}(a)–(b) become indistinguishable in the presence of a dissipative environment.

{\it Circuit QED Hamiltonian framework ---}  
Being interested in the quantum properties of the circuits in Fig.~\ref{fig:circuits}, it is convenient to describe them using the Hamiltonian formalism \cite{vool2017introduction}. For the details of the derivations and properties of Hamiltonians see \cite{supp}. We first consider the capacitively-terminated \textit{charge} circuit (Fig.~\ref{fig:circuits}(a)), which was also analyzed in \cite{giacomelli2024emergent}. Its Hamiltonian can be expressed in different \textit{gauges}, and in the so-called charge gauge is given by  
\begin{equation}\label{eq:charge-hamiltonian}
	\begin{split}
		\mathcal{\hat H}^{ch} =& 4E_C (\hat N  -\nu)^2- E_J\cos\hat\varphi\\
		&+\sum_{k=1}^{N_m}\h\w_k\hat{a}_k^\dag \hat{a}_k+i (\hat N-\nu)\sum_{k=1}^{N_m} g_k(\hat a_k^\dag-\hat a_k),
	\end{split}
\end{equation}
where $\hat{a}_k$ are bosonic annihilation operators for the transmission line modes, satisfying $[\hat{a}_k, \hat{a}_k^\dagger] = 1$, and $\hat{N}$ and $\hat{\varphi}$ are the dimensionless charge and phase operators, satisfying the canonical commutation relation $[\hat{\varphi}, \hat{N}] = i$.
The periodicity of the Hamiltonian in $\hat\varphi$ reflects the fact that the charge in the upper part of the circuit can only change in integer increments by tunneling through the junction. Here, $\nu$ represents a gate bias charge, which can be interpreted as a Bloch quasi-charge for the full system \cite{giacomelli2024emergent}. The mode frequencies $\w_k$ and couplings $g_k$ are computed numerically for given values of the impedance $Z$, the cutoff frequency $\omega_c$, and the low-energy mode spacing $\Delta$. At low frequencies, these couplings produce an Ohmic bath with scaling behavior $g_k^2/\hbar\w_k\sim 2\hbar\Delta Z/R_q$.

The Hamiltonian of the short-ended \textit{flux} circuit (Fig.\ref{fig:circuits}(b)) is more naturally formulated in the flux gauge, yielding 
\begin{equation}\label{eq:dual-hamiltonian}
	\begin{split}
		\mathcal{\hat H}^{fl} = &\frac{\Phi_0^2}{4}\frac{\bar\w_c}{\bar Z}\hat\varphi^2 + 4\bar E_C \hat N^2 - \bar E_J\cos(\hat\varphi+2\pi \phi) \\ &+\sum_{k=1}^{N_m}\h\Omega_k\hat{a}_k^\dag \hat{a}_k+  \hat\varphi\sum_{k=1}^{N_m} f_k(\hat a_k+\hat a_k^\dag)\, .
	\end{split}
\end{equation}
As before, $[\hat{a}_k, \hat{a}_k^\dagger] = 1$ and $[\hat{\varphi}, \hat{N}] = i$. Here, $\phi$ denotes an external flux bias applied to the entire array, expressed in units of the flux quantum $\Phi_0 = h/2e$.
 As in the charge circuit, the mode frequencies $\Omega_k$ and couplings $f_k$ are determined numerically and also give rise to an Ohmic bath at low frequencies, with $f_k^2/\hbar\Omega_k\sim (2\pi^2)^{-1}\hbar\bar\Delta R_q/\bar Z$, exhibiting an inverse (dual) dependence on the impedance.  
Importantly, if this Hamiltonian were expressed in the charge gauge, an inductive term would still be present \cite{supp}, unlike in Eq.~\eqref{eq:charge-hamiltonian}, breaking the symmetry under phase translations. This reflects the absence of charge quantization in the circuit. Instead, flux through the array remains quantized, as it can only change by integer amounts due to phase slips at the junction.

The presence of the charging and Josephson terms in Eq.~\eqref{eq:dual-hamiltonian} prevents it from being exactly dual to Eq.~\eqref{eq:charge-hamiltonian}, meaning the two Hamiltonians do not maintain the same form under the exchange of fluxes and charges. This duality of the Hamiltonians is usually restored only in the limit where charging effects are treated perturbatively, corresponding to the large $E_J/E_C$ regime. This is the core assumption behind the commonly invoked single-band approximation:  
\begin{equation}\label{eq:one-band-approx}
	4E_C \hat N^2 - E_J\cos\hat\varphi \simeq U_0\cos(2\pi\hat N),
\end{equation}
where $U_0$ is the bandwidth (phase slip amplitude), exponentially small in $E_J/E_C$ \cite{koch2009charging}. This simplification serves as the foundation for the approximate duality, that effectively treats the junction as a phase-slip element. However, in this work, we avoid such an approximation to explore an exact duality.

\begin{figure}[t]
  \centering
  \includegraphics[width=\columnwidth]{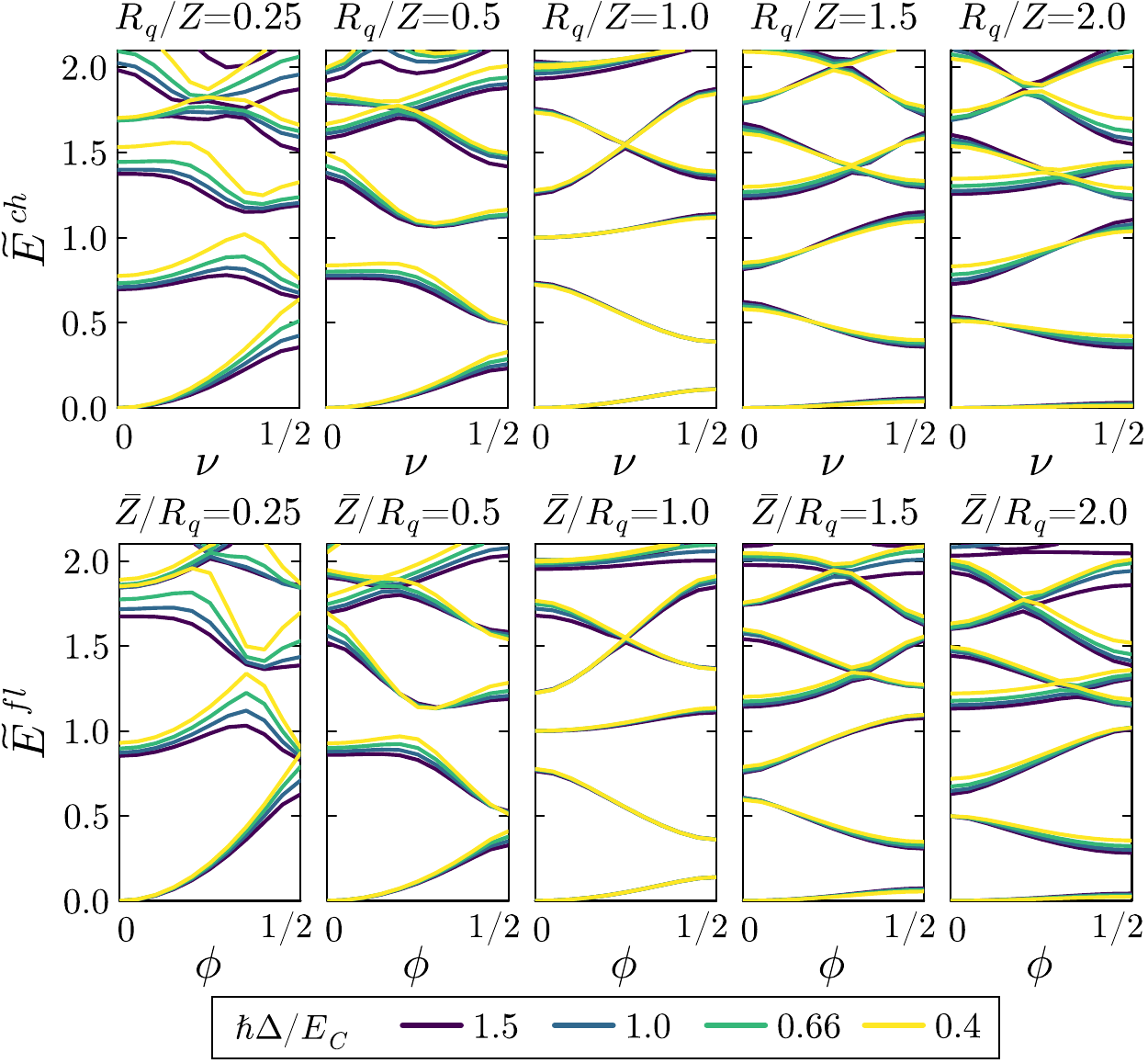}
\caption{Top: Rescaled energy bands of the charge circuit [Fig.~1(a)] vs. gate charge $\nu$, shown for different system lengths (mode spacing $\Delta$) and various $R_q/Z$ values. Here $\tld{E}^{ch}(Z,\nu)=E^{ch}(Z,\nu)/E^{ch}_3(R_q,0)$. 
Bottom: Rescaled energy bands of the flux circuit [Fig.~1(b)] vs. normalized external flux $\phi$, for the same inverse ratios $\bar{Z}/R_q$. Here $\tld{E}^{fl}(\bar{Z},\phi)=E^{fl}(\bar Z,\phi)/E^{fl}_3(R_q,0)$. 
Parameters: $E_J = E_C$, $\bar{E}_J = E_J$, $\bar{E}_C = E_C$, $\bar{\omega}_c = \omega_c = 4E_C/\hbar$. The four transmission lines considered have lengths of 5, 7, 10, and 16 $LC$ unit cells, respectively. }
  \label{fig:bands}
\end{figure}

{\it Exact solution and dual Brillouin zones ---}
For a finite-length transmission line, the Hamiltonians in Eqs.~\eqref{eq:charge-hamiltonian} and \eqref{eq:dual-hamiltonian} can be numerically solved by using exact diagonalization. The case of the capacitively terminated circuit was analyzed in \cite{giacomelli2024emergent}. Here, we extend this approach to the more complex case of the dual circuit. Specifically, the first three terms in the Hamiltonian can be interpreted as describing a flux-biased fluxonium atom with an impedance-dependent inductance. Notably, at large impedances, the inductive energy scale can become very small, leading to a dense fluxonium spectrum that requires accounting for a large number of energy levels. This contrasts sharply with Eq.~\eqref{eq:charge-hamiltonian}, where only a few levels of the transmon-like Hamiltonian, which is $2\pi$ phase periodic, are needed for each charge bias $\nu$.

Unlike in \cite{giacomelli2024emergent}, we perform numerical diagonalization in the \textit{polaron frame}, where linear interaction terms in Eqs.~\eqref{eq:charge-hamiltonian} and \eqref{eq:dual-hamiltonian} are absorbed via a displacement of photonic operators. This leads to faster convergence with a smaller photonic Fock space, accelerating the solution of Eq.~\eqref{eq:charge-hamiltonian} and allowing Eq.~\eqref{eq:dual-hamiltonian} to be solved for longer systems (for more details, see \cite{supp}). As previously discussed, the full spectrum of Eq.~\eqref{eq:charge-hamiltonian} forms energy bands over the Brillouin zone $\nu \in [-1/2, 1/2]$ due to phase periodicity. Although Eq.~\eqref{eq:dual-hamiltonian} is not explicitly periodic in charge, its spectrum is periodic in external flux $\phi$. Thus, one can restrict $\phi \in [-1/2, 1/2]$ and interpret the spectrum as composed of bands in a \textit{dual Brillouin zone}. This can also be understood from the periodicity of Eq. \eqref{eq:dual-hamiltonian} in the quasi-charge $\nu$ \cite{supp}.

To demonstrate the exact duality summarized in Fig.~\ref{fig:circuits}(c), we follow the path that led us to uncover it. We begin with the two circuits in Fig.~\ref{fig:circuits}(a)-(b) with impedances $Z$ and $\bar{Z} = R_q^2/Z$, while keeping other parameters equal: $\bar{E}_J = \bar{E}_C = E_J = E_C$, $\bar{\Delta} = \Delta$, and $\bar{\omega}_c = \omega_c$. Fig.~\ref{fig:bands} shows the corresponding energy band dispersions. Results are plotted for five impedance values $Z$ and corresponding $\bar{Z}$ (different panels), and for four line lengths—i.e., four values of mode spacing $\Delta$ (different line colors). 
For comparison, energies are rescaled by the gap between the first and third bands at $Z = R_q$, a quantity that rapidly converges to $\hbar \Delta$ as the system size increases.
The top row shows the charge-dependent bands of Fig.~\ref{fig:circuits}(a); the bottom row shows the flux-dependent bands of Fig.~\ref{fig:circuits}(b). Notably, both circuits exhibit a length-independent low-energy spectrum at $Z = R_q$, confirming criticality in both cases. Moreover, the energy bands closely match, with anticrossings aligned in the dual Brillouin zones.

Importantly, the energy spectra of the two circuits display opposite dependencies on the ratio $E_J/E_C$, as illustrated in Fig.~\ref{fig:mobilities}(a)-(b) (supplementary impedance values are provided in \cite{supp}). Specifically, the charge-dependent energy bands of the charge circuit become increasingly flat as $E_J/E_C$ grows, whereas the flux-dependent bands of the flux circuit develop pronounced dispersions and narrow anticrossings. In other words, a strongly interacting regime in one circuit corresponds to a weakly interacting regime in the other, and vice versa. While the well-known charge-flux duality \cite{schmid1983diffusion,bulgadaev1984phase,guinea1985diffusion,weiss2012quantum} is typically invoked in the asymptotic limits $E_J \ll E_C$ and $E_J \gg E_C$, here we observe the emergence of a self-duality that extends across the full range of $E_J/E_C$, even in regimes where the commonly used single-band approximation of the junction~\eqref{eq:one-band-approx} is no longer applicable. 
By matching the energy scales of the two transmission lines—namely, $\bar{E}_C = E_C$, $\bar{\Delta} = \Delta$, and $\bar{\omega}_c = \omega_c$—we observe an exact low-energy duality, meaning that the low-energy band structures are identical. This is achieved by choosing $\phi = \nu$, $\bar{Z} = R_q^2 / Z$, and, crucially, by transforming the Josephson energy according to an impedance- and length-dependent function, $\bar{E}_J = \mathcal{F}_{\omega_c, Z, \Delta}(E_J)$, as we demonstrate in the following.

\begin{figure}[t!]
  \centering
  \includegraphics[width=\columnwidth]{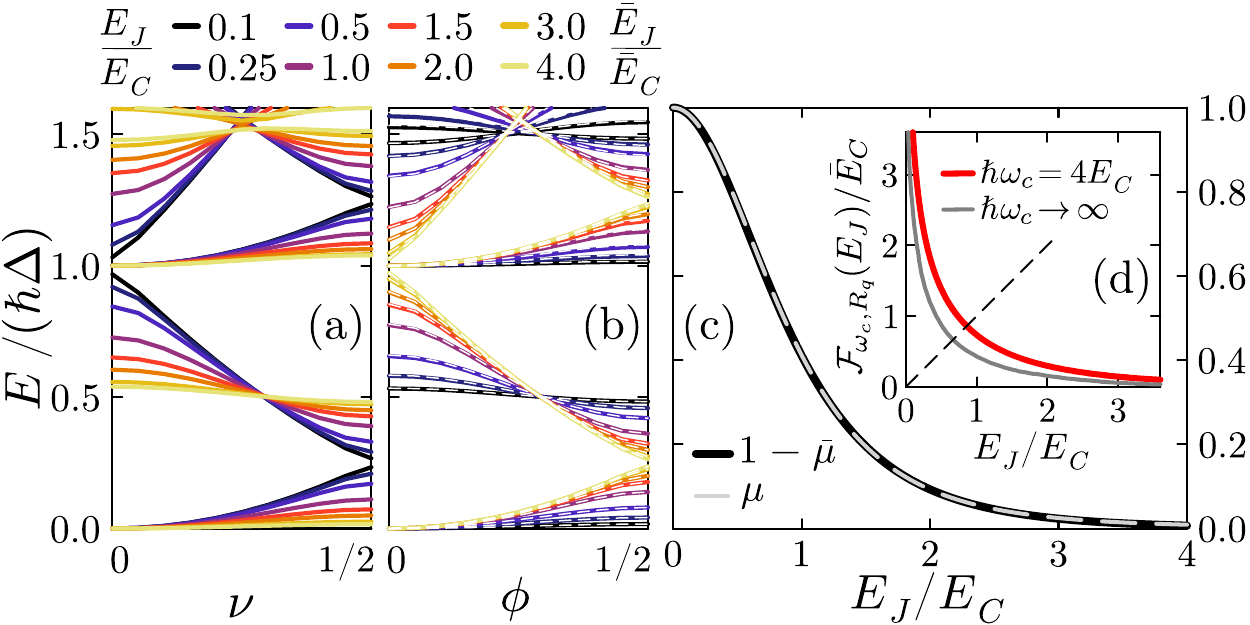}
\caption{(a) Band spectrum of the charge circuit at the critical point $Z = R_q$ for various $E_J/E_C$ values (different colors). (b) Same for the flux circuit at $\bar{Z} = R_q$. Thin dashed white lines show the analytical prediction \eqref{eq:cft-spectrum}. Parameters: $\hbar \Delta = 0.66 E_C$, $\hbar \omega_c = 4 E_C$. (c) Mobility at criticality vs. $E_J/E_C$ for both circuits. For the flux case, we plot $1 - \bar{\mu}$ to highlight the exact duality between mobilities. (d) Critical self-dual transformation $\mathcal{F}_{\omega_c, R_q}(E_J)$ preserving the band shape. The thick red line is from numerics at $\hbar \omega_c = 4 E_C$; the thin grey line interpolates the $\omega_c \to \infty$ results from \cite{lukyanov2007resistively}. Intersections with the unit-slope line give the duality center $E_J^*$, where $\mathcal{F}_{\omega_c, R_q}(E_J^*) = E_J^*$ and the spectra coincide.}

  \label{fig:mobilities}
\end{figure}

{\it Critical line and the center of the self-duality ---} 
The observed independence of the rescaled low-energy spectrum at $Z = R_q$ signals the emergence of conformal invariance and indicates that our finite-size numerical results are already remarkably close to the thermodynamic limit. Interestingly, the critical spectra exhibit a simple structure. In Ref.~\cite{hasselfield2006critical}, an analytical expression was derived based on the boundary sine-Gordon model under the assumption of conformal invariance. The resulting band structure depends on a single parameter, the \textit{mobility} $\mu \in [0,1]$, and takes the form:
\begin{equation}\label{eq:cft-spectrum}
	\frac{E_\mathrm{CFT}(\xi)}{\hbar\Delta} = \left( \lambda(\mu,\xi) + n \right)^2 + p,
\end{equation}
where $\lambda(\mu,\xi) = \cos^{-1}\left( \sqrt{\mu} \cos(2\pi\xi) \right) / 2\pi$, and $\xi \in [-1/2, 1/2]$ spans the Brillouin zone.

This conformal field theory (CFT) prediction was shown to match well with the numerical critical spectrum of the charge circuit (with $\xi = \nu$), and the numerical data was used to extract the non-analytical relation between $E_J/E_C$ and the mobility $\mu$ \cite{paris2024resilience}. We now extend this comparison to the dual bands, for which $\xi = \phi$. For a given value of $\bar{E}_J/\bar{E}_C$, we use Eq.~\eqref{eq:cft-spectrum} with $n = 0$ and $p = 0$ to fit the mobility $\bar{\mu}$ from the lowest numerical band, and then compare the full band structure. The resulting CFT prediction is plotted as thin dashed white lines in Fig.~\ref{fig:mobilities}(b), overlaid on the numerical spectrum. As shown, the agreement is excellent across the first few bands and for all values of $E_J/E_C$. 

In Fig.~\ref{fig:mobilities}(c), we plot the fitted critical-point mobilities $\mu$ and $\bar{\mu}$ as functions of the ratio $E_J/E_C$, considering the case where $\bar{E}_J = E_J$ and $\bar{E}_C = E_C$. The mobility $\mu$ of the charge circuit (thin dashed grey line) is found to satisfy the relation $\mu = 1 - \bar{\mu}$, where $\bar{\mu}$ is the mobility of the flux circuit (thick black line). This relation can be proven within the standard formulation of duality~\cite{schmid1983diffusion,bulgadaev1984phase}, and we show that it holds for all values of $E_J/E_C$. From these data, we extract the self-duality transformation along the critical line—namely, the transformation $\bar{E}_J = \mathcal{F}_{\omega_c, R_q}(E_J)$ for which the band structure of the flux circuit becomes identical to that of the charge circuit. This transformation is implicitly defined by the condition $\mu(E_J) \equiv \bar{\mu}(\mathcal{F}_{\omega_c, R_q}(E_J))$, and is plotted in Fig.~\ref{fig:mobilities}(d) (thick red line). Note that the exact form of $\mathcal{F}_{\omega_c, R_q}$ depends on the cutoff frequency $\omega_c$, which sets the ultraviolet cutoff of the environment and directly influences the mobility~\cite{paris2024resilience}.  For comparison, we also plot the corresponding function extracted from an interpolation of the Monte Carlo results reported in Ref.~\cite{lukyanov2007resistively}, which were obtained in the limit of infinite $\omega_c$. Interestingly, this exact self-duality also implies the existence of a precise \textit{center of the self-duality}, defined by the condition $\mathcal{F}_{\omega_c,R_q}(E_J^*) = E_J^*$, where the two circuits exhibit identical energy bands. This point corresponds to the value of $E_J/E_C$ for which the mobility satisfies $\mu = 0.5$.
Remarkably, this self-dual point lies in the regime $E_J < E_C$, with the limiting value in the $\omega_c \to \infty$ case given by $E_J^*/E_C \simeq 0.66$. 

\paragraph{\it Duality across the phase diagram ---}  
At the critical point $Z = R_q$, the low-energy spectra of both circuits with a finite transmission line have already converged to their infinite-length limits, where the two systems become indistinguishable. In this regime, the observed spectral duality is also a \emph{self-duality} of the infinite-length model. Away from this critical line, the circuits continue to exhibit a precise low-energy \emph{spectral duality} at any finite size: their low-energy bands share the same shape under the transformation illustrated in Fig.~\ref{fig:circuits}(c), which relates the two Josephson energies via $\bar{E}_J = \mathcal{F}_{\omega_c, Z, \Delta}(E_J)$, accounting for impedance and system length. This correspondence arises because the positions of band anticrossings in the Brillouin zone depend solely on the impedance $Z$, as shown in Fig.~\ref{fig:bands}, while the parameters $E_J$, $\omega_c$, and $\Delta $ control the band widths, that can again characterized by single parameters $\mu$ and $\bar\mu$, plotted with two heatmaps in Fig.\ref{fig:circuits}(e) (further quantitative analysis is provided in~\cite{supp}). In summary, the duality transformations connecting different quadrants of the phase diagram—depicted with arrows in Fig.~\ref{fig:circuits}(e)—originate from this exact finite-size low-energy spectral duality, which persists throughout the scaling towards the limit $\Delta \to 0$. Note that the low-energy duality uncovered in this work is expected to hold for energies much smaller than the plasma frequency of the Josephson tunneling junctions: indeed, at this energy scale the higher bands of the flux circuit begin to play a significant role.

{\it Conclusions ---}  
We have demonstrated that the self-duality of the resistively shunted Josephson tunnel junction arises from two distinct Caldeira-Leggett representations of the environment, corresponding to the circuits in Fig.~\ref{fig:circuits}(a)-(b), which differ in their low-frequency behavior. Rather than assuming duality, we solved both models exactly across the full range of $E_J/E_C$, revealing that the approximate self-duality widely used in the literature becomes exact along the entire critical line. This implies that the phase transition is independent of $E_J/E_C$, and the identical continuous spectra emerge in opposite interaction regimes for the two circuits. Furthermore, the dual Hamiltonian~\eqref{eq:dual-hamiltonian}, under the one-band approximation~\eqref{eq:one-band-approx}, describes a phase-slip junction or superconducting nanowire coupled to a shorted transmission line—i.e., the exact dual of Fig.~\ref{fig:circuits}(a). Our results therefore apply directly to nanowire-based implementations. Finally, this framework opens new directions for generalizations to more complex superconducting circuits. The concept of exact duality could serve as a powerful guide in identifying critical behavior in broader classes of superconducting–insulating transitions. 

From an experimental perspective, using transmission lines instead of resistors prevents unwanted heating in the circuit. The system can also be probed via its environmental photonic modes, as demonstrated in \cite{kuzmin2025observation} for the small-$E_J/E_C$ regime. This approach can be extended to larger values of $E_J/E_C$ to test the duality directly. Further details are provided in the End Matter, where we present numerical results for the low-frequency photon spectroscopy of both circuits—an alternative to conventional IV-characterization measurements. In practice, experiments are expected to reproduce the theoretical predictions near the critical impedance and close to the self-dual point, where the rates of Cooper pair tunneling and of quantum phase slips are similar, and the system can equilibrate within the experimental timescale.

\acknowledgements
This work was supported by the French project TRIANGLE (ANR-20-CE47-0011); a grant from the French government managed by the National Research Agency (ANR) under the France 2030 program (reference ANR-24-RRII-0001); HPC resources from the French GENCI–IDRIS computing facility; and support from DARPA under grant HR0011-24-2-0346 and ARO under grant W911NF-23-1-0051. We also acknowledge insightful discussions with Christophe Mora and Federico Borletto.

\bibliography{biblio.bib}

\onecolumngrid

\vspace{1cm}

\begin{center}
{\textbf{End Matter}}
\end{center}

\twocolumngrid

{\it Photon spectroscopy —}
A key advantage of studying the Schmid–Bulgadaev transition using finite-length transmission-line resonators is that the environmental photonic modes can be directly observed. This approach was employed in \cite{kuzmin2025observation} to characterize the transition as $E_J/E_C$ was tuned by applying a magnetic field through a SQUID. The experiment, performed at small values of $E_J/E_C \sim 0.1$, revealed frequency shifts of the modes that were successfully reproduced at low energies by exact diagonalization in \cite{giacomelli2024emergent}.

To probe the duality, we extend this calculation to the flux circuit and to larger values of $E_J/E_C$. From the eigenstates obtained by exact diagonalization at zero external bias ($\nu = 0$ for the charge circuit and $\phi = 0$ for its dual), we compute the photonic spectral function
\begin{equation}
    D(\omega)=\sum_n\textstyle{\frac{\gamma^2}{\gamma^2+(\hbar\omega-E_n-E_G)^2}}\sum_{k=1}^{N_m}|\bra{G}\hat a_k^\dagger\ket{E_n}|^2,
\end{equation}
where $\ket{E_n}$ and $E_n$ denote the eigenstates and eigenenergies, $\ket{G}$ and $E_G$ are the ground-state and its energy, and $\gamma$ is a phenomenological linewidth.

In Fig.~\ref{fig:spectroscopy}, this quantity is shown for both circuits, for three impedance values and across the full relevant range of $E_J/E_C$. The first row displays the results for the charge circuit. At small $E_J/E_C$, the photonic modes exhibit frequency shifts of opposite sign on the two sides of the transition, consistent with the observations of \cite{kuzmin2025observation}. For larger values of $E_J/E_C$, secondary peaks appear and eventually merge.

\begin{figure}[t!]
  \centering
  \includegraphics[width=\columnwidth]{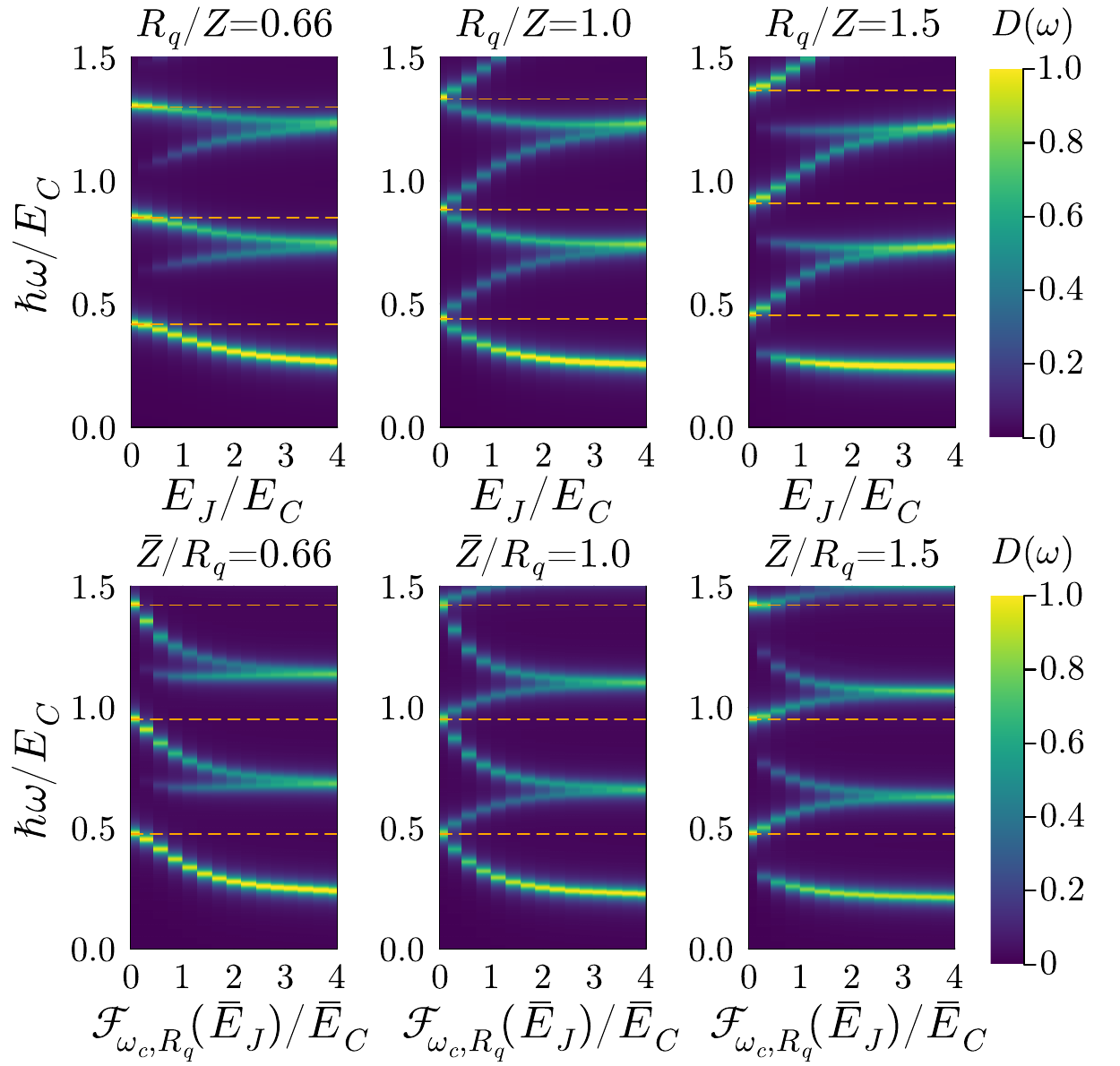}
\caption{Photonic spectral function versus  the Josephson-to-charging energy ratio for the two circuits, shown for three impedance values. Top row: results for the charge circuit. Bottom row: results for the flux circuit at the corresponding dual impedances, plotted versus the Josephson energy transformed according to the critical duality relation. Thin dashed orange lines indicate the bare mode frequencies entering Hamiltonians \eqref{eq:charge-hamiltonian} and \eqref{eq:dual-hamiltonian}. Parameters: $\bar{\omega}_c = \omega_c = 10E_C/\hbar$, $\bar{\Delta} = \Delta = 0.5E_C/\hbar$.}

  \label{fig:spectroscopy}
\end{figure}

In the second row of Fig.~\ref{fig:spectroscopy}, we show the results for the flux circuit, obtained using the same environmental parameters. When the horizontal axis is transformed according to the critical duality mapping $\mathcal{F}_{\omega_c,R_q}$, the two circuits exhibit identical spectral features at the corresponding dual impedances. For comparison, Fig.~\ref{fig:spectroscopy-dual-notransf} displays the $\bar{E}_J/\bar{E}_C$ dependence of the spectral function without applying this transformation.

This analysis illustrates how the duality between the two circuits can be directly probed through the spectroscopy of their photonic modes. In particular, for the flux circuit, as $\bar{E}_J/\bar{E}_C$ is decreased from large values, the frequencies of the modes decrease for $\bar Z<R_q$ and increases for $\bar Z>R_q$. This is followed by the emergence of additional peaks that mirror those of the charge circuit at the corresponding dual parameters. This behavior provides a clear experimental signature of the charge–flux duality across the transition, which can be measured for example through linear reflection spectroscopy as in \cite{kuzmin2025observation}.

As discussed above, the exact duality is expected to hold at energies lower than the charging and plasma energies of the junctions. In Fig.~\ref{fig:spectroscopy}, due to intrinsic numerical limitations, the separation between these energy scales and the mode spacing is not fully achieved, and residual finite-size effects persist. In experimental realizations, however, the free spectral range can be made much smaller, enabling a test of the duality over a broader set of modes.

\begin{figure}[t!]
  \centering
  \includegraphics[width=\columnwidth]{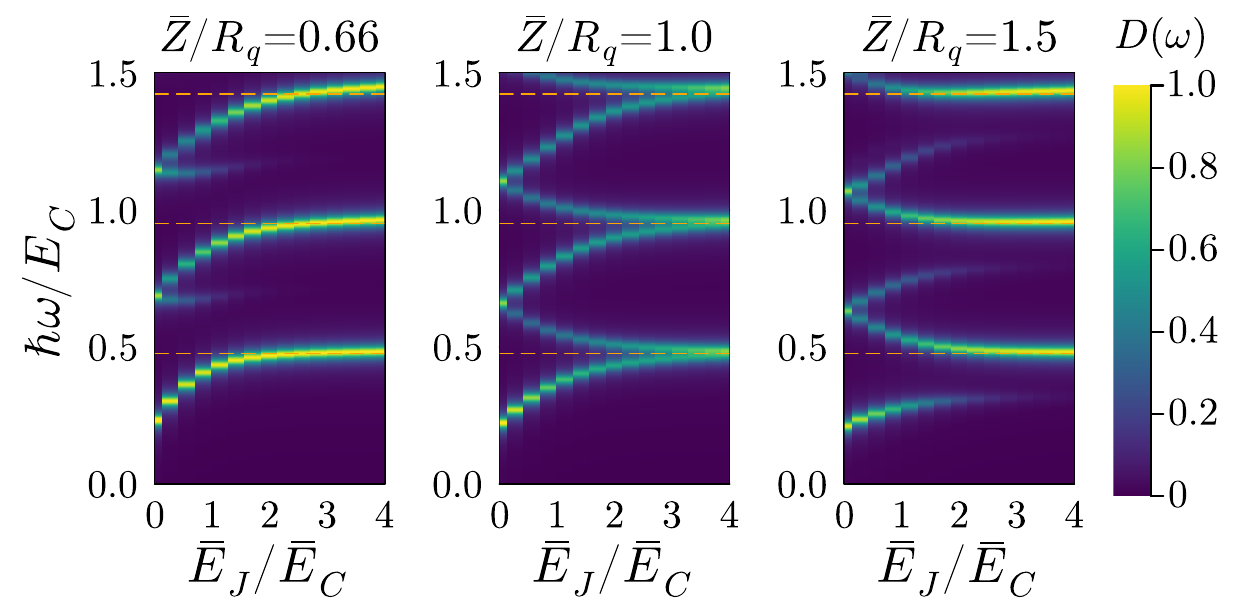}
\caption{Photonic spectral function for the flux circuit, plotted as a function of the original Josephson energy, complementing the results shown in the second panel. Thin dashed orange lines indicate the bare mode frequencies entering Hamiltonians \eqref{eq:charge-hamiltonian} and \eqref{eq:dual-hamiltonian}. Parameters: $\bar{\omega}_c = \omega_c = 10E_C/\hbar$, $\bar{\Delta} = \Delta = 0.5E_C/\hbar$.} \label{bare_EJ}
  \label{fig:spectroscopy-dual-notransf}
\end{figure}


\newpage
\clearpage

\title{{\bf Supplementary Material for the article:}\\ ``Exact Duality at Low Energy in a Josephson Tunnel Junction \\ Coupled to a Transmission Line"}
\setcounter{page}{1}
\setcounter{equation}{0}
\setcounter{figure}{0}
\renewcommand{\theequation}{S.\arabic{equation}}
\renewcommand{\thefigure}{S.\arabic{figure}}
\pagestyle{empty}
\date{\today}
\maketitle
\onecolumngrid

\section{Hamiltonian derivation and gauges}

The Hamiltonian description of the circuits can be derived starting from their Lagrangian written in terms of the capacitance and inductance matrices $\mathbf{C}$ and $\mathbf{\Gamma}$ and of the vector of fluxes $\bm{\varphi}^T=(\varphi,\varphi_1,\dots,\varphi_N)$, one for each active node of the circuit.
\begin{equation}\label{eq:LC-lagrangian}
	\mathcal{L}=\frac{1}{2}\frac{\h^2}{8E_C}\dot{\bm{\varphi}}^T \mathbf{C}\dot{\bm\varphi} - \frac{1}{2}\frac{\h^2}{8E_C}\bm\varphi^T\mathbf{\Gamma}\bm\varphi+ E_J\cos\varphi
    \hspace{.2cm}\mbox{ with }
    \mathbf{C}=\frac{2E_C}{e^2}
 {\footnotesize\arraycolsep=0.3\arraycolsep\ensuremath{
	\begin{pmatrix}
	C_J &  & &  & \\
	 	& C &  & & \\
		&  & & \ddots  & \\
	 & & &  & C
	\end{pmatrix}
	}},
	\hspace{.2cm}
	\mathbf{\Gamma}=\frac{2E_C}{e^2}\frac{1}{L}
	{\footnotesize\arraycolsep=0.3\arraycolsep\ensuremath{
	\begin{pmatrix}
	1 	& -1	&		 & &\\
	-1	& 2 	& -1 	 & & \\
	 	& -1 	& \ddots & \ddots \\
	 	&  		& \ddots & \ddots & -1 \\
	 	& 		& 		 & -1	 & B
	\end{pmatrix}
	}}.
\end{equation}
The last entry on the diagonal of $\mathbf{\Gamma}$ is $B=1$ for the open boundary condition (Fig.1(a)) and $B=2$ for the short one (Fig.1(b)). The charge bias in the open-ended circuit can be included as an extra term in the Lagrangian $-\nu\dot\varphi$, while the flux bias of the short-ended circuit can be included as a shift in the argument of the cosine $\cos(\varphi+\Phi)$.

It is convenient to express the Lagrangian in terms of the eigenmodes of the uncoupled transmission line, which are obtained with a tranformation acting only on the fluxes $\bm{\widetilde \varphi}=(\varphi_1,\dots,\varphi_N)$ that simultaneously diagonalizes the submatrices $\mathbf{\tld C}$ and $\mathbf{\tld\Gamma}$ of $\mathbf{C}$ and $\mathbf{\Gamma}$ obtained by excluding the first row and column. In particular, we apply a linear transformation $\bm{\widetilde\varphi}=\mathbf{P}\bm{\widetilde\phi}$ such that $\mathbf{\tld\Gamma P}=\mathbf{\tld C P}\bm{\Omega}^2$, with $\bm{\Omega}$ a diagonal matrices containing the eigenmodes frequencies. We also choose the normalization $\mathbf{P^T\tld C P}=\bm{I}$. The resulting expression of the Lagrangian is
\begin{equation}\label{eq:sm-lagrangian}
	\mathcal{L}=\frac{1}{2}\frac{\hbar^2}{8E_C}\sum_{i=1}^N \left( \dot{\phi}_i^2 -(\Omega_{i}^B)^2 \phi_i^2 \right)+\frac{1}{2}\frac{\hbar^2}{8E_C} \dot{\varphi}^2-\frac{1}{2}\frac{\Phi_0^2}{L} \varphi^2+\frac{\Phi_0^2}{L} \varphi \sum_{i=1}^N P^B_{1i}\phi_i+E_J \cos\varphi,
\end{equation}
where $\Phi_0=\hbar/2e$ is the flux quantum and the superscript $B$ indicates the two boundary conditions, that change the frequencies and the eigenmodes. We can pass to the Hamiltonian formalism by defining the conjugate momenta $N=\hbar^{-1}\partial \mathcal{L}/\partial \dot\varphi$ and $n_i=\hbar^{-1}\partial \mathcal{L}/\partial \dot\phi_i$ and we can canonically quantize these variables and introduce creation and annihilation operators through $\hat\phi_i=\sqrt{4E_C/\hbar\Omega_i^B}(\hat a_i+\hat a_i^\dagger)$ and $\hat n_i=-i\sqrt{\hbar\Omega_i^B/16E_C}(\hat a_i-\hat a_i^\dagger)$. We obtain the \textit{flux gauge} Hamiltonian
\begin{equation}\label{eq:sm-flux-gauge}
	\mathcal{\hat H}_B^{fl}=4E_C \hat N^2 -E_J\cos\hat\varphi +\frac{1}{2}\frac{\Phi_0^2}{L}\hat\varphi^2 + \sum_i \hbar\Omega_i^B\hat a_i^\dagger \hat a_i - \hat\varphi \sum_i \underbrace{\frac{\Phi_0^2}{L}\sqrt{\frac{4E_C}{\hbar\Omega^B_i}}P^B_{1i}}_{f^B_i}(\hat a_i+\hat a_i^\dagger).
\end{equation}
Where we introduced the couplings $f_i^B$, that essentially depend on the amplitude of the transmission line modes at the first node of the array. This representation for $B=2$ is the one we use in equation (2) of the main text for the flux circuit.

For $B=1$ an analytical solution for the frequencies and the couplings can be obtained \cite{borletto2024circuit}, while for $B=2$ such a solution is not available for a finite number of modes. Also considered the following manipulations of the Hamiltonian, for simplicity here we resort to numerical computation of these quantities.

At low frequencies the mode spacing is the same for both boundary conditions, but the two sets of eigenfrequencies are shifted, namely $\Omega_n^1\sim(n-1/2)\Delta$ and $\Omega_n^2\sim n\Delta$. Also, both couplings have the same dependence on the frequency and on the array characteristic impedance $Z=\sqrt{L/C}$ at low frequencies
\begin{equation}\label{eq:sm-flux-couplings}
	f^B_n\sim \frac{\sqrt{\hbar\Delta}}{\sqrt{2}\pi}\sqrt{\frac{R_q}{Z}}\sqrt{\hbar\Omega^B_i}.
\end{equation}
A crucial difference between the two arrays is the fact that, for the capacitively-terminated one ($B=1$) the following \textit{sum rule} holds
\begin{equation}\label{eq:sm-sumrule}
	\sum_i\frac{(f^1_i)^2}{\hbar\Omega^1_i}=\frac{\Phi_0^2}{2L}.
\end{equation}
This does not hold for $B=2$ and reflects the absence of a closed inductive loop in the open-ended circuit, as will be clearer in the following. This equality makes it so that the $\hat\varphi^2$ term in the Hamiltonian is a so-called Caldeira-Leggett \textit{counterterm}, that is usually artificially introduced to model dissipation from an unknown source that should not modify the \textit{bare} potential \cite{caldeira1983quantum,weiss2012quantum} (in other words, to complete the square given by the inductive terms of the modes and the coupling term). Here it is naturally present, and implies that we can think of the open-ended circuit as a representation of the usual Caldeira-Leggett model.

We named Hamiltonian \eqref{eq:sm-flux-gauge} flux gauge Hamiltonian due to the appearence of the coupling through the flux variables. This is analogous to the dipole gauge in cavity QED. It is convenient to consider the gauge in which the coupling is through the charge variables, that can be obtained through the unitary transformation
\begin{equation}\label{eq:sm-gauge-transformation}
	\mathcal{U}_{fl\to ch}=\exp\left(-\hat\varphi\sum_i\frac{f^B_i}{\hbar\omega^B_i} (\hat a_i^\dag-\hat a_i)\right).
\end{equation}
This gives the following \textit{charge gauge} Hamiltonian
\begin{equation}\label{eq:sm-charge-gauge}
	\mathcal{\hat H}_B^{ch} = 4E_C \left(\hat N  + i\sum_i\frac{f^B_i}{\hbar\Omega^B_i}(\hat a_i^\dag - \hat a_i)\right)^2- E_J\cos\hat\varphi+ \left(\frac{\Phi_0^2}{2L} - \sum_i\frac{(f_i^B)^2}{\hbar\Omega_i^B}\right)\hat\varphi^2+ \sum_{i}\h\Omega^B_i\hat{a}_i^\dag \hat{a}_i.
\end{equation}
The first term is reminiscent of minimal coupling with the electromagnetic field, and this gauge is analogous to the Coulomb gauge of cavity QED.

Notice that, while for the open-ended case $B=1$ the sum rule \eqref{eq:sm-sumrule} assures that the inductive term vanishes, for the short-ended case $B=2$ this is not true, and an inductive scale given by the inductance of the whole array remains. This crucially influences the symmetries of the Hamiltonian. In fact, for the open-ended case the Hamiltonian is invariant under translations of $\hat\varphi$ of multiples of $2\pi$, while for the short-ended case it is not. This symmetry is given by the fact that the charge in the upper part of the open-ended circuit can only change by integer amounts. Notice also that the transformation \eqref{eq:sm-gauge-transformation} mixes the degrees of freedom, so that in the charge gauge $\hat\varphi$ is no longer the phase difference of the junction.

A further useful manipulation can be made on Hamiltonian \eqref{eq:sm-charge-gauge} to get rid of the diamagnetic-like term coming from the square of the environmental charge in the first term. This is a quadratic term in the environmental operators and can be brought to a diagonal form with a multi-mode Bogoliubov transformation, that gives new frequencies of the independent modes $\omega_i$ and new couplings $g_i$. This leaves the charge gauge Hamiltonian in the shape given in equation (1) of the main text for $B=1$.

There is no closed analytical formula for the eigenfrequencies and the couplings after the Bogoliubov transformation, we hence determine them numerically. However, to get an intuition we can use the approximate expression derived in \cite{masuki2022absence} for mode frequencies much lower than the plasma frequency. In particular
\begin{equation}
	g_i\sim \sqrt{2\hbar\Delta}\sqrt{\frac{Z}{R_q}}\sqrt{\frac{\h\w_i}{1+\left(\frac{\pi\h\w_i}{4E_C}\frac{Z}{R_q}\right)}}.
\end{equation}
At large frequencies $\w_i\gg \frac{1}{ZC_J}$ (given by the $RC$ time) the couplings behave as the ones appearing in \eqref{eq:sm-charge-gauge}, i.e. $g_i\sim f_i/\sqrt{\omega_i}$. At low frequency instead, $g_i\sim \sqrt{2\h\Delta}\sqrt{\h\w_i}\sqrt{Z/R_q}$. In our language, this dependence, \textit{dual} to \eqref{eq:sm-flux-couplings} comes from the Bogoliubov transformation. Notice that, to get to equation (1) in the main text, one can directly perform a simultaneous diagonalization of the full $\bm{C}$ and $\bm{\Gamma}$ matrices \eqref{eq:sm-lagrangian}, as was done in \cite{giacomelli2024emergent}.


\begin{figure*}[t]
  \centering
  \includegraphics[width=0.9\columnwidth]{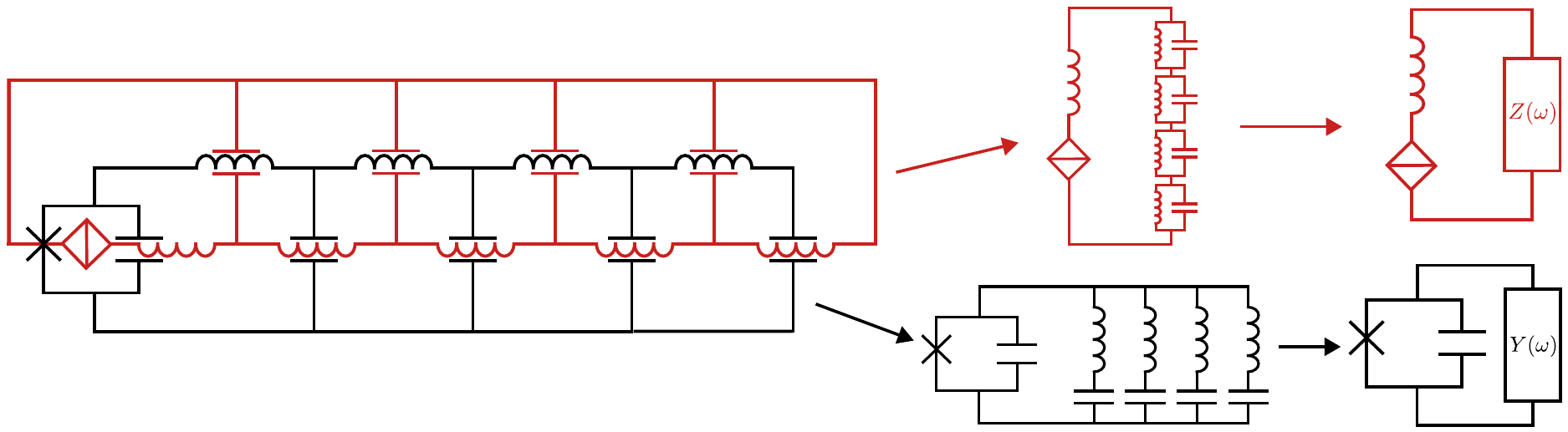}
  \caption{Graphical construction of the dual circuit. The capacitively-terminated transmission line is dual to the inductively terminated one, while the parallel between the tunneling element and the capacitor is dual to an inductor and a phase slip element in series. The two transmission lines are realizations of the Caldeira-Leggett for an impedance and an admittance.}
  \label{fig:circuit-duality}
\end{figure*}

\section{Circuit duality and Caldeira-Leggett models}

The dual of an electrical circuit is the circuit that satisfies the same equations with currents and voltages interchanged. In other words, the mesh equations of a circuit have the same form of the nodal equations of the other one. The dual circuit can be constructed graphically by exchanging nodes and meshes, and by substituting each element with the one with the dual (inverse) impedance (see for example \cite{hayt2012engineering}).

In the leftmost diagram of Fig.\ref{fig:circuit-duality} we perform such a construction for the capacitively-terminated lumped-element transmission line. The dual turns out to be an inductively-terminated transmission line. In substituting the dual elements one should maintain the same \textit{numerical} values of the impedances, so that the characteristic impedance of the dual (shorted) line is the inverse of the original one. The two circuits also have the poles and zeros of the impedance interchanged \cite{pozar2021microwave}, corresponding to the shifted eigenfrequencies discussed in the previous section.

Beyond the \textit{classical} circuit elements, we also introduced a phase slip element, that is the dual of the Josephson tunnel junction \cite{mooij2006superconducting}. If the Josephson tunnel junction behaves as a non-linear inductor, the phase slip element behaves as a non-linear capacitor, while charge tunneling is dual to phase slips. The phase slip element behaviour is naturally obtained for a Josephson tunnel junction if the one-band approximation is made, as discussed in the main text, which amounts to describe phase slips involving well-defined phase states.

From the Hamiltonian framework point of view the circuit duality is a charge-phase duality, in the sense that the Hamiltonian of the dual circuit has the same form with exchanged flux and charge variables and opposite characteristic impedance. This guarantees exact dual physics also in the quantum regime. 

In the middle diagrams of Fig.\ref{fig:circuit-duality} we show an alternative representation of the two transmission lines as a series of LC oscillators or as a parallel of them (in these the inductances and the capacitances are not all equal), with the key difference of the impedance at zero frequency, which is zero in the top circuit and infinite in the bottom one. When the number of modes is taken to infinity these two diagrams are the Caldeira-Leggett representations respectively of an impedance and an admittance \cite{devoret2021does}, as indicated in the rightmost diagrams of Fig.\ref{fig:circuit-duality}. In the standard treatment both circuits, even though they are purely reactive, develop a real part of the impedance in the limit of infinite number of modes and zero mode spacing, and can hence be used to model dissipation in a Hamiltonian framework. 

It is important to notice that, in the case considered in the main text, we do not take the dual circuit of the Josephson junction coupled to a capacitively-terminated transmission line. Instead, we couple the same Josephson junction to two different Caldeira-Leggett representations of the environment, with the same characteristic impedance. Even though the two representations of the environment have different conserved quantities, in the \textit{thermodynamic} limit they become indistinguishable, i.e. the different boundary conditions become immaterial. The two Hamiltonian are hence not dual to one another, and the exact duality we find is a non-trivial property of the low-energy spectrum in large enough systems.


\section{Bloch representation}
In the main text we plot the spectra of the two considered circuits as bands living in a Brillouin zone. The \textit{quasimomenta} that span these Brillouin zones have a clear physical meaning in terms of the conserved quantities in the two circuits: the \textit{quasicharge} $\nu$ is a gate charge biasing the charge island present in the upper part of the capacitively terminated circuit, while the \textit{quasiflux} $\Phi$ is a flux biasing the loop present in the inductively terminated one. These can however also be introduced formally with an application of Bloch theorem to a periodic Hamiltonian.

For what concerns the capacitive-ended circuit, as we already discussed, the discrete periodicity in $\varphi$ of the charge gauge Hamiltonian is a consequence of charge discretization. Different values of the gate charge $\nu$ are hence separate problems, but it is important to consider the dependence on this parameter because when the system is perturbed, for example by a current bias, the different quasicharges become coupled.

In \cite{giacomelli2024emergent}, Bloch theorem was applied to the Hamiltonian (1) in the main text, i.e. the invariance of the Hamiltonian under the phase translations $\mathcal{\hat U}_1^p=e^{i(2\pi p)\hat N}$, with $p$ an integer, implies that eigenstates have the form $\braket{\varphi}{\nu,s}=e^{i\nu\varphi}u_\nu^s(\varphi)$, where $s$ is the band index and the Brillouin zone is $\nu\in[-1/2,1/2]$.

In the inductively terminated circuit the application of Bloch theorem is less immediate. In fact, the Hamiltonian is not invariant under charge translations, as it would be with the one band approximation. Still, it is invariant under quasicharge translations. To see this, one can rewrite Hamiltonian (2) in the main text in terms of Bloch states for the uncoupled junction (that is the second and the third terms)
\begin{equation}
	\mathcal{\hat H}_2^{fl} = \frac{E_L}{2}\hat\varphi^2 + \sum_{\nu,s}\varepsilon_s(\nu)\ket{\nu,s}\bra{\nu,s}+\sum_{k=1}^{N_m}\h\Omega_k\hat{a}_k^\dag \hat{a}_k+  \hat\varphi\sum_{k=1}^{N_m} f_k(\hat a_k+\hat a_k^\dag),
\end{equation}
where $\varepsilon_s(\nu)$ are the eigenenergies. Just as the charge operator can be written as $\hat N=\hat\nu+\mathcal{\hat P}$, with $\hat\nu$ diagonal in the bands, the phase operator can be split as $\hat\varphi=\hat\phi + \mathcal{\hat Q}$. Here $\hat \phi$ is the \textit{Bravais lattice operator}, is diagonal in the bands and is the canonical conjugate to the quasicharge $[\hat\phi,\hat\nu]=i$ (see e.g. \cite{balian1989relation}). This means that the operator $\mathcal{\hat U}_2^p=e^{ip\hat\phi}$ is a quasicharge translation of an integer amount $p$.

The second term is trivially invariant under such an operation since the spectrum repeats in the neighboring Brillouin zones. The first term is less obviously invariant, since $[\hat\Phi,\mathcal{\hat Q}]\neq 0$. However, when expressed in terms of Bloch functions, it is diagonal in $\nu$
\begin{equation}
	\mathcal{\hat Q}=\sum_{\nu,s,r}\mathcal{Q}_{sr}(\nu)\ket{\nu,s}\bra{\nu,r}.
\end{equation}
Since $\mathcal{\hat U}_2\ket{\nu,s}=\ket{\nu+1,s}$, and $\ket{\nu+1,s}$ differs from $\ket{\nu,s}$ for a phase, also this operator is left invariant by the quasicharge translation.

One can hence take advantage of this periodicity and use Bloch theorem to build a \textit{dual Brillouin zone}, spanned by the \textit{quasi-flux} $\phi\in [-1/2,1/2]$. The problem can be solved separately for each value of $\phi$ with the reduced Hamiltonian
\begin{equation}
	\mathcal{\hat H}_\phi^{fl} = \frac{E_L}{2}(\hat\varphi-2\pi\phi)^2 + \sum_{\nu,s}\varepsilon_s(\nu)\ket{\nu,s}\bra{\nu,s}+\sum_{k=1}^{N_m}\h\Omega_k\hat{a}_k^\dag \hat{a}_k+  (\hat\varphi-2\pi\phi)\sum_{k=1}^{N_m} f_k(\hat a_k+\hat a_k^\dag).
\end{equation}
One can get to Hamiltonian (2) in the main text, with the external flux inside the cosine, with the unitary transformation $e^{i2\pi\phi\hat N}$.

Notice that, even though the application of Bloch theorem in this dual case is nontrivial, and only becomes standard when the one band approximation is taken, the periodicity of the spectrum in the external flux $\phi$ is to be expected on physical grounds.


\section{The polaron frame}

As also discussed in \cite{giacomelli2024emergent,paris2024resilience} for the capacitively terminated circuit, it is useful to also write the Hamiltonians in a different \textit{frame}, that we will call the polaron frame. This is in analogy with the Lee-Low-Pines transformation used for polaron problems, and was more recently discussed in a cavity QED setting in \cite{ashida2021cavity}. Here we give a slightly different discussion compared to \cite{giacomelli2024emergent} and connect directly with the polaron transformation.

In our case, we consider the following unitary on Hamiltonian (1)
\begin{equation}
	\mathcal{\hat U}_p=\exp\left(i\hat N\sum_k\frac{g_k}{\hbar\w_k}(\hat a_k+\hat a_k^\dagger)\right).
\end{equation}
This gives us
\begin{equation}\label{eq:sm-polaron-frame-charge}
	\mathcal{\hat U}_p \mathcal{\hat H}_1^{ch}\mathcal{\hat U}_p^\dagger= \underbrace{\left(4E_C-\sum_k\frac{g_k^2}{\h\w_k}\right)}_{4\tld E_C}\hat N^2 - E_J\cos\left(\hat\varphi+\sum_k\frac{g_k}{\h\w_k}(\hat a_k+\hat a_k^\dagger)\right) + \sum_k\h\w_k\hat a_k^\dagger \hat a_k,
\end{equation}
where we defined the \textit{renormalized charging energy} $\tld E_C$. This form of the Hamiltonian is also a starting point to obtain the boundary sine Gordon model \cite{morel2021double,paris2024resilience}. The main effect of this transformation is to get rid of the linear coupling between the junction and the modes.

Let us consider the matrix elements of the cosine term in \eqref{eq:sm-polaron-frame-charge} between two eigenstates of $\hat N$ and Fock states for all the (polaron frame) photons labeled by a vector of occupation numbers $\bra{M,\bm{m}}\mathcal{\hat U}_p \cos\hat\varphi\ \mathcal{\hat U}_p^\dagger\ket{N,\bm{n}}$. It is easier to consider the unitaries as acting on the states, giving a $N$ dependent displacement $\hat D(iN\frac{g_k}{\h\w_k})$ of each photonic mode, where $\hat D(\alpha)=e^{\alpha\hat a^\dagger-\alpha^*\hat a}$. Let us denominate the displaced number states by 
\begin{equation}\label{eq:polaron-basis}
	\ket{N,\bm{\tld n}}=\ket{N}\bigotimes_k\hat D\left(iN\frac{g_k}{\h\w_k}\right)\ket{n_k}. 
\end{equation}
One needs hence to consider
\begin{equation}\label{eq:sm-cosine-jump}
	\begin{split}
	\bra{M,\bm{\tld m}} E_J\cos\hat\varphi\ \ket{N,\bm{\tld n}}&= 	\bra{M,\bm{\tld m}} \frac{E_J}{2}\sum_{N'}(\ket{N'}\bra{N'+1}+\ket{N'+1}\bra{N'})\ \ket{N,\bm{\tld n}}\\
	&=\frac{E_J}{2}\delta_{M,N-1}\braket{N-1,\bm{\tld m}}{N,\bm{\tld n}} + \frac{E_J}{2}\delta_{M,N+1}\braket{N+1,\bm{\tld m}}{N,\bm{\tld n}}.
	\end{split}
\end{equation}
We can now use the overlap between two displaced number states \cite{cahill1969ordered}
\begin{equation}
	\bra{m}\hat D^\dagger(\beta)\hat D(\alpha)\ket{n}=e^{\frac{1}{2}(\alpha\beta^*-\alpha^*\beta)}e^{-\frac{1}{2}|\alpha-\beta|^2}\times \sqrt{\frac{n!}{m!}}(\alpha-\beta)^{(m-n)}L_n^{(m-n)}(|\alpha-\beta|^2),
\end{equation}
where $m>n$ and $L_n^{(m)}$ are the associated Laguerre polynomials. The matrix element \eqref{eq:sm-cosine-jump} then becomes
\begin{equation}\label{eq:sm-cos-polaron}
	\bra{M,\bm{\tld m}} E_J\cos\hat\varphi\ \ket{N,\bm{\tld n}}=
	\frac{1}{2}\underbrace{E_J e^{-\frac{1}{2}\sum_{k}\frac{g_k^2}{\h^2\w_k^2}}}_{\tld{E}_J} \left(\delta_{M,N-1}+(-1)^{|\bm{m}-\bm{n}|}\delta_{M,N+1}\right) \prod_{k=1}^{N_m} \sqrt{\frac{n_k!}{m_k!}}\left(\frac{g_k}{\h\w_k}\right)^{m_k-n_k}L_{n_k}^{(m_k-n_k)}\left(\frac{g_k^2}{\h^2\w_k^2}\right),
\end{equation}
where we introduced the \textit{renormalized Josephson energy} $\tld E_J$. This factor reduces the magnitude of all the matrix elements.

\section{One band approximation and elastic renormalization}

\begin{figure}[t!]
  \centering
  \includegraphics[width=0.5\columnwidth]{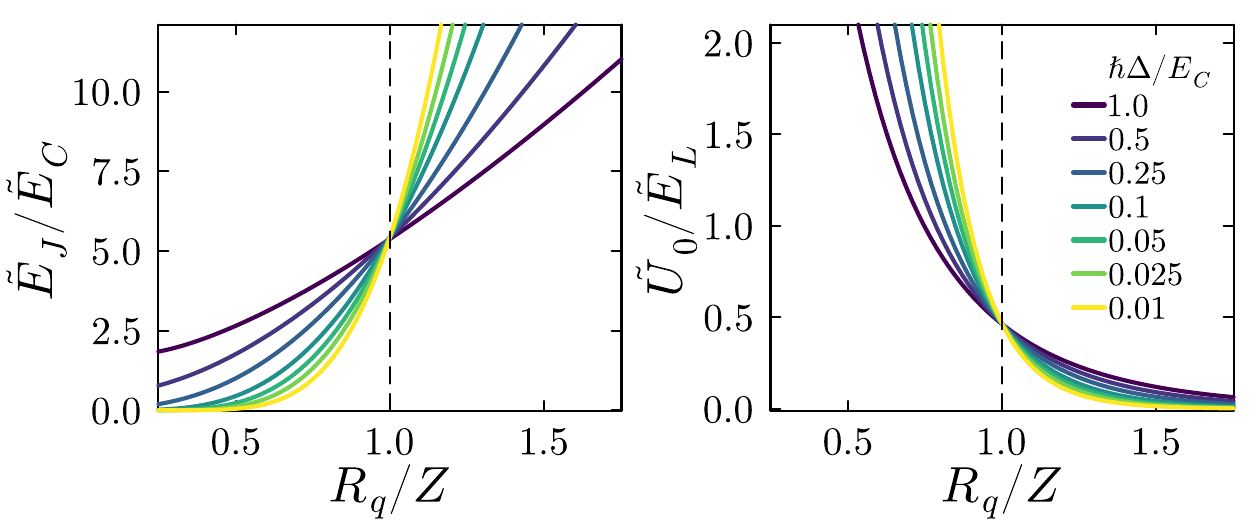}
  \caption{Ratio of the elastically renormalized quantities in the two circuits for different system sizes. The left panel shows the direct case, while the right panel depicts the dual case in the one-band approximation. Here, $E_J/E_C=4$ and $\omega_p=4E_C$.}
  \label{fig:elastic}
\end{figure}

As a first approximation one can consider Hamiltonian \eqref{eq:sm-polaron-frame-charge} in the subspace with zero (polaron frame) photons. The Hamiltonian is hence only characterized by the renormalized charging energy and the renormalized Josephson energy, that account for the \textit{elastic} dressing of the junction. This renormalization corresponds to the adiabatic elimination of the environmental modes and depends only on the frequencies and the couplings. In Fig.\ref{fig:elastic}(a) we show the ratio of these two quantities as a function of impedance for different system sizes (this calculation was already performed in \cite{giacomelli2024emergent} and is reported here for the comparison with the dual case). 

We can apply the same reasoning to the flux gauge Hamiltonian (2) once the single-band approximation (3) is taken. With this approximation, one is neglecting inter-band tunneling and identifying the quasicharge with the full charge. This is equivalent to considering a tight-binding description with hopping between neighboring wells \cite{guinea1985diffusion}, or by an adiabatic approach \cite{paris2024resilience}. In other terms, with approximation \eqref{eq:one-band-approx} one is treating the junction as a superconducting nanowire phase-slip junction, which is the exact dual of the Josephson tunnel junction \cite{mooij2006superconducting} and for which the flux through the loop is quantized.

With this approximation $\mathcal{\hat H}_2^{fl}$ becomes periodic in $\hat N$ and of the same shape as $\mathcal{\hat H}_1^{ch}$, with charges and fluxes interchanged and with the following correspondences
\begin{equation}
	4E_C\leftrightarrow \frac{E_L}{2};\hsp E_J\leftrightarrow V_0;\hsp \frac{R_q}{Z}\to\frac{Z}{R_q}.
\end{equation}

Notice that in this case the polaron transformation that removes the linear coupling term is the gauge transformation \eqref{eq:sm-gauge-transformation}. In this case one obtains a renormalized inductive energy, dual to the renormalized capacitive energy. While the renormalized capacitance can be thought of as the (very large) total capacitance of capacitively shorted array, the renormalized inductance is the very large inductance of the loop present in the inductively terminated array. The cosine in the charge can be written in terms of flux eigenstates as
\begin{equation}
	4E_C\hat N^2-E_J\cos\hat\varphi\simeq U_0\cos(2\pi\hat N)=\sum_n \Big( \ket{2\pi n}\bra{2\pi(n+1)} + \ket{2\pi (n+1)}\bra{2\pi n} \Big),
\end{equation}
so that a suppression of $U_0$ dual to the one identified in \eqref{eq:sm-cos-polaron} can be derived. Summing up one has
\begin{equation}
	\widetilde E_L=E_L-2\sum_k\frac{f_k^2}{\hbar\Omega_k};\hsp \tld U_0=U_0\exp\left(-\frac{4\pi^2}{2}\sum_k\frac{f_k^2}{\hbar^2\Omega_k^2}\right).
\end{equation}
The ratio of this quantity is shown in Fig.\ref{fig:elastic}(b) for different sizes of the transmission line. 

Both circuits display a size-independence of the ratio at $Z=R_q$, value that separates the two \textit{phases} in which one of the two renormalized scales dominate. This approximation becomes exact in the $E_J\to 0$ or $U_0\to 0$ limit, in which the polaron frame photons become non-interacting, and can be expected to hold perturbatively beyond that. Moreover, the scale invariance of the exact spectra, shown for both circuits in Fig.2 of the main text, indicates that this predicted scale invariance is exact. 

Away from the critical point, while for the direct case the energy scale of the nonlinearity $\widetilde E_J$ dominates at small impedances $Z<R_q$, the opposite is true for the corresponding scale $\widetilde U_0$ in the dual case, that dominates at large impedances $Z>R_q$. The localized/delocalized regimes are hence interchanged, as expected.


\section{Exact diagonalization in the polaron frame}
The effect of the redefinition of the photons in the polaron frame is to reabsorb the \textit{classical} contribution to the number of photons and to hence reduce the number of photons needed to describe the exact states. This is convenient for numerical diagonalization because one can work with a lower cutoff on the Fock space of the photons and hence work with a smaller Hamiltonian. For best results, one also needs to pay attention to the choice of the gauge in which the starting Hamiltonian is represented, as it can be crucial once the numerical truncation of the Hilbert space is performed.

For the charge gauge Hamiltonian (1) the diagonalization procedure is the same used in \cite{paris2024resilience}, that is we use states \eqref{eq:polaron-basis} as a basis to represent the Hamiltonian, with the analytical matrix elements of the cosine operator \eqref{eq:sm-cos-polaron}. For each bias charge $\nu$, we first find $N_{bands}$ eigenstates of the isolated junction and express the charge eigenstates $\ket{N}$ as a superposition of them. The photonic Hilbert space is truncated to contain all the states with an energy smaller than $E_{cut}$. Numerical convergence is obtained by increasing $N_{bands}$ and $E_{cut}$; see \cite{paris2024resilience} for an analysis of the convergence.

For the flux gauge Hamiltonian (2) instead, we use as as basis the states
\begin{equation}\label{eq:polaron-basis-flux}
	\ket{\varphi,\bm{\tld n}}=\ket{\varphi}\bigotimes_k\hat D\left(\varphi\frac{f_k}{\h\W_k}\right)\ket{n_k},
\end{equation}
where $\ket{\varphi}$ are phase eigenstates. These are obtained as superpositions of the first $N_{lev}$ eigenstates of the fluxonium-like Hamiltonian given by the first three terms of (2). The Hamiltonian is then constructed and diagonalized analogously to the previous case. Notice that in this case $N_{lev}$ must be larger than $N_{bands}$ in the previous case, as the eigenstates can also be very close in energy if the effective inductive scale is small.



\begin{figure}[t!]
  \centering
  \includegraphics[width=\columnwidth]{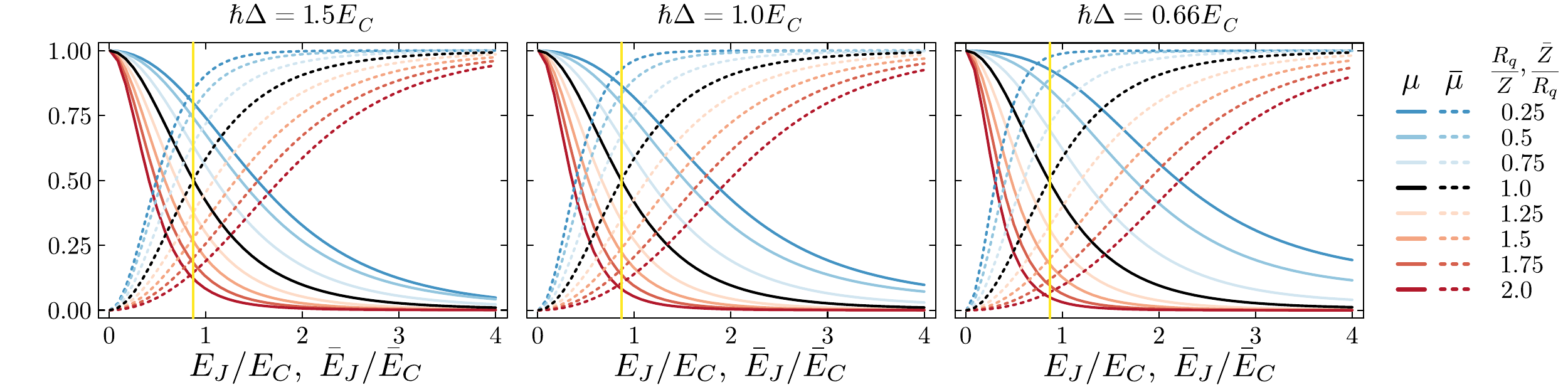}
  \caption{Fitting parameters $\mu$ and $\bar{\mu}$ for the spectra of the two circuits, shown for various impedance values and system lengths. The finite-size duality transformation can be extracted by identifying points where $\mu = \bar{\mu}$ and $R_q/Z = \bar{Z}/R_q$. The parameters are computed with $\hbar \omega_p = 4E_C$. The vertical yellow line indicates the center of self-duality at $E_J^*$.}
  \label{fig:duality-other-Z}
\end{figure}

\section{Duality Transformation Away from the Critical Line}

In the main text, we identified the transformation $\bar{E}_J = \mathcal{F}_{\omega_p, R_q}(E_J)$ that realizes exact self-duality along the critical line $Z = R_q$. Since our spectra at criticality already represent the thermodynamic limit, this transformation is independent of the level spacing $\Delta$. While this confirms that the critical line is independent of the ratio $E_J/E_C$, the same transformation cannot be directly applied away from the critical line. In this regime, the spectra depend on the system length, and the two circuits remain distinguishable due to finite-size effects, preventing direct extraction of a self-duality mapping from the numerical data.

Nonetheless, as illustrated in Fig.~2 of the main text, the two circuits—although not dual at the circuit level—exhibit low-energy spectral duality. One can therefore study the emergence of self-duality away from the critical line. Crucially, a well-defined duality transformation still exists in this case, but it acquires an explicit dependence on the system length: $\bar{E}_J = \mathcal{F}_{\omega_c, Z, \Delta}(E_J)$. This duality can be quantified by two key observations:  
(i) the positions of the anticrossings in the spectrum depend solely on the impedance, while their widths are controlled by $E_J$, $\Delta$ and $\omega_c$;  
(ii) the lowest energy band can always be fitted using the conformal field theory prediction (4), provided the bandwidth is treated as a fitting parameter. Thus, the low-energy spectrum is fully characterized by the impedance and a fitted parameter $\mu$.

In Fig.~\ref{fig:duality-other-Z}, we plot the extracted fitting parameters: $\mu$ for the charge circuit (solid lines) and $\bar{\mu}$ for the flux circuit (dashed lines), for several impedance values (indicated by color). For a given color, the data correspond to the same numerical value of $R_q/Z$ in the charge circuit and of $\bar{Z}/R_q$ in the flux circuit. Three different system lengths are shown. As expected, the black curves for $Z = \bar{Z} = R_q$ remain unchanged with system length, while the other curves vary with $\Delta$. The dual values of $E_J$ and $\bar{E}_J$ for fixed impedance are obtained by drawing a horizontal line and identifying the intersections with the same-color solid and dashed curves. We thus conclude that low-energy duality persists at finite size for arbitrary impedance values, with a duality transformation that depends on the system length. In the thermodynamic limit, the entire phase diagram becomes self-dual, with the transformation $\bar{E}_J = \lim_{\Delta \to 0} \mathcal{F}_{\omega_c, Z, \Delta}(E_J)$ governing the Josephson energy mapping.

\begin{figure}[t!]
  \centering
  \includegraphics[width=0.66\columnwidth]{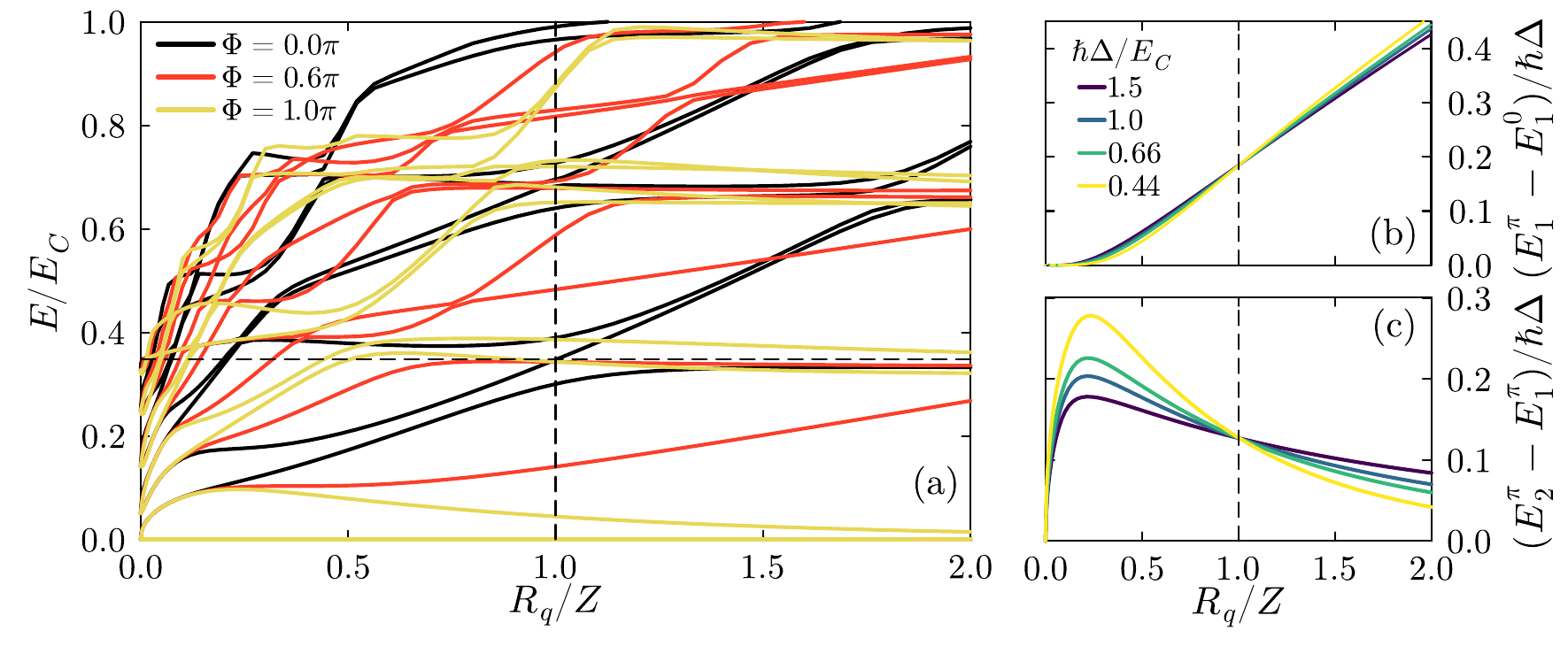}
  \caption{(a) Impedance dependence of the spectrum of the dual circuit for three values of the external biasing flux. Here $E_J=2E_C$, $\omega_p=2E_C$ and $\hbar\Delta=0.44E_C$. (b) Difference in the ground state energies at $\Phi=0$ and $\Phi=\pi$ for different system sizes. (c) Gap above the ground state for $\Phi=\pi$ for different system sizes.}
  \label{fig:short-spectrum}
\end{figure}

\section{Additional exact diagonalization results}

In Fig.~\ref{fig:short-spectrum}(a), we show the spectra of the inductively terminated circuit in a different way, by plotting the impedance dependence of the spectra for three values of the external biasing flux $\Phi$. From the perspective of the effective fluxonium atom, at $\Phi=0$, the potential is even and has a single absolute minimum. The full spectrum in this case (black line) exhibits an avoided crossing centered at the critical point, occurring at the energy of the first photonic mode. For $\Phi=\pi$ (yellow line), the potential remains symmetric but now features two degenerate minima. The corresponding spectrum contains two levels below the first photonic frequency (horizontal dashed line), with the energy gap decreasing at lower impedances.

The low-lying spectrum maintains the same qualitative shape across different system sizes, shifting to lower energies proportional to $\hbar\Delta$ (as highlighted in Fig.2). In Fig.\ref{fig:short-spectrum}(b), we plot the impedance dependence of the energy difference between the ground states at $\phi=\pi$ and $\phi=0$ (the width of the first dual band). When rescaled by the free spectral range $\Delta$, this quantity grows linearly around $Z=R_q$ for various system sizes. All curves intersect at $Z=R_q$, with the slope increasing for smaller free spectral ranges (corresponding to larger systems). In panel (c), we plot the energy gap between the ground state and the first excited state at $\Phi=\pi$ (band gap at the edge of the dual Brillouin zone). The rescaled curves again intersect at $Z=R_q$, with larger systems exhibiting an increasingly gapless behavior for $R_q/Z>1$.

Notably, in the infinite-size limit, the system remains gapless, and the flux dependence of the spectrum vanishes. This is expected, as the transmission line asymptotically behaves as a resistive element, eliminating flux dependence. However, the scale invariance observed at $Z=R_q$ in Fig.\ref{fig:short-spectrum}(b)-(c) indicates the presence of a critical point, with the system flowing in opposite directions away from it. This confirms that the approximate result in Fig.\ref{fig:elastic}(b) holds at an exact level. 

\begin{figure*}[t] \centering \includegraphics[width=\textwidth]{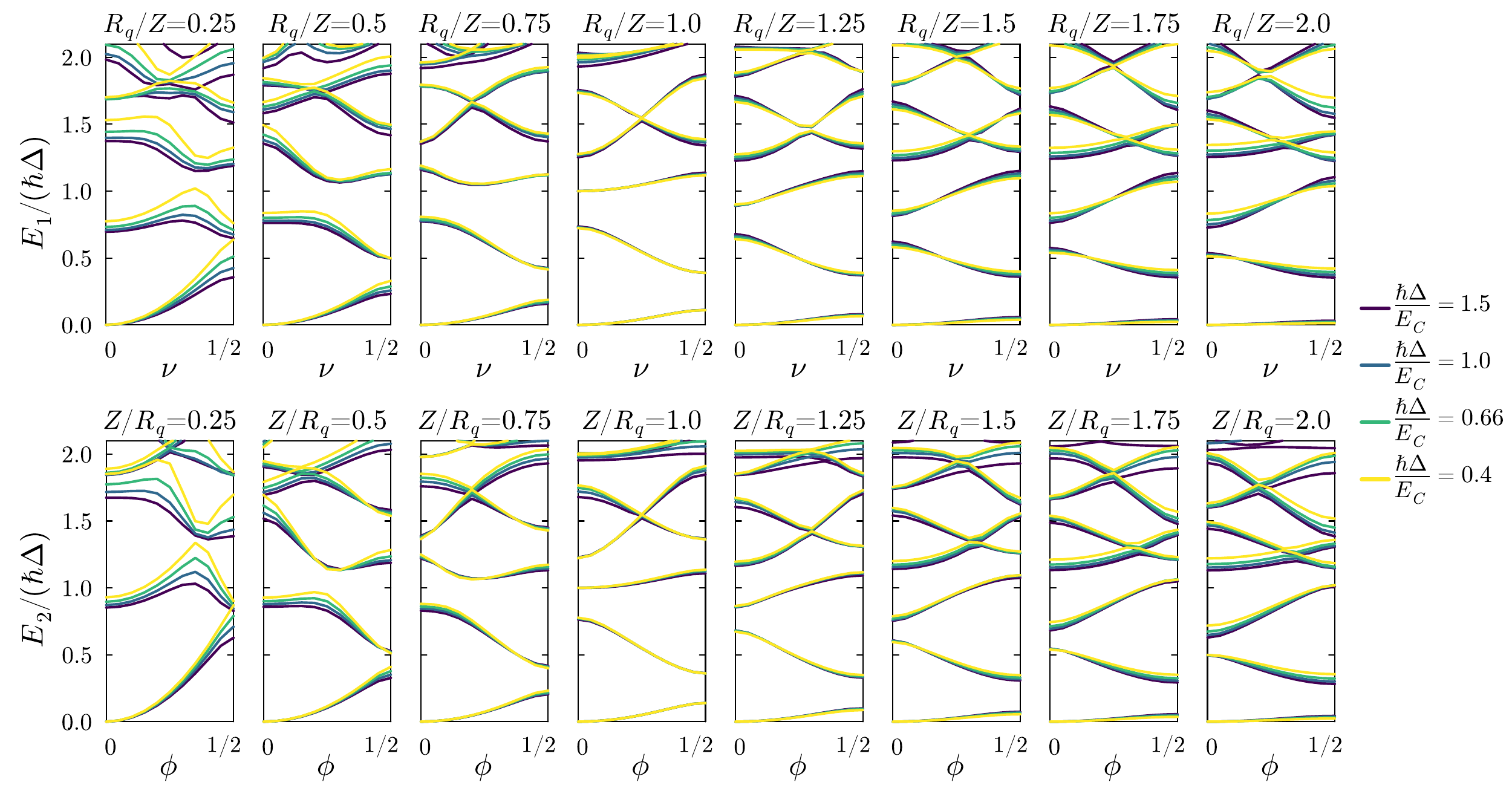} \caption{Top: Bands of the open-ended circuit as a function of the gate charge $\nu$ for different system sizes and impedance values.
Bottom: Bands of the dual circuit as a function of the external flux $\Phi$, with impedance values inverted relative to the top panels.
Parameters: $E_J=E_C$ and $\hbar\omega_p=4E_C$.} \label{fig:extended_bands_EJ1} \end{figure*}

We also present supplementary numerical data for the band structures plotted in Fig.2 and 3 of the main text. In Fig.\ref{fig:extended_bands_EJ1}, we display band spectra for additional impedance values corresponding to the parameters used in Fig.~2. One can observe a qualitative duality in the size dependence of the band spectra across all impedance values.

\begin{figure*}[t] \centering \includegraphics[width=\columnwidth]{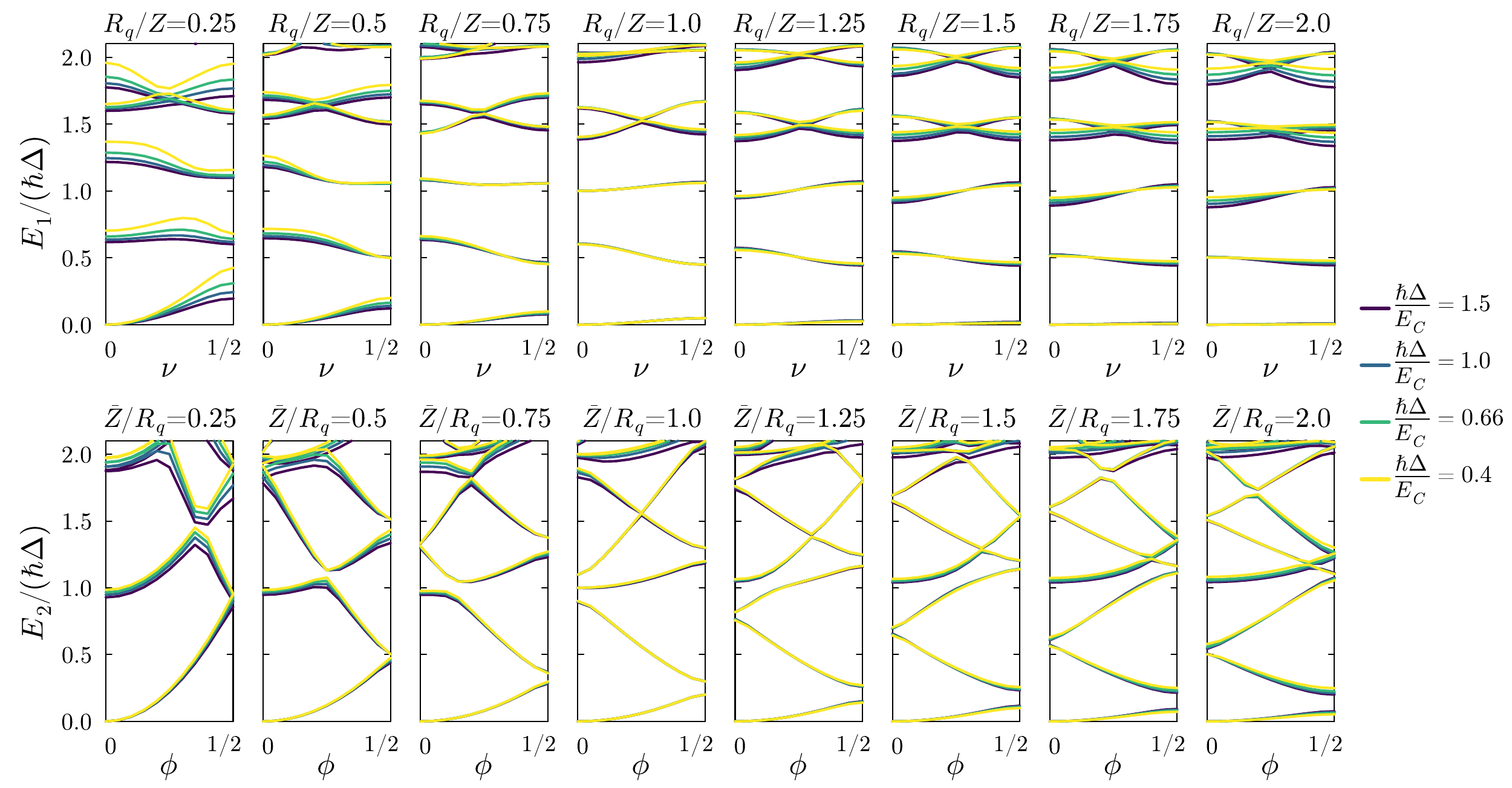} \caption{Direct (upper panels) and dual (lower panels) bands for $E_J=2E_C$ and $\omega_p=4E_C$.
Compared to the spectra for $E_J=E_C$ shown in Fig.~\ref{fig:extended_bands_EJ1}, the anticrossings in the direct bands become wider, while those in the dual bands become narrower.} \label{fig:extended_bands_EJ2} \end{figure*}

In Fig.~\ref{fig:extended_bands_EJ2}, we present analogous plots for a larger ratio of $E_J/E_C=2$. Notably, the direct band spectra exhibit stronger interactions, characterized by wider anticrossings, while the dual band spectra become more weakly interacting. These plots illustrate the opposite behavior of the two spectra with respect to variations in $E_J/E_C$ across several impedance values, mirroring the behavior observed in Fig.~3 of the main text.

\begin{figure}[t] \centering 
\includegraphics[width=0.5\textwidth]{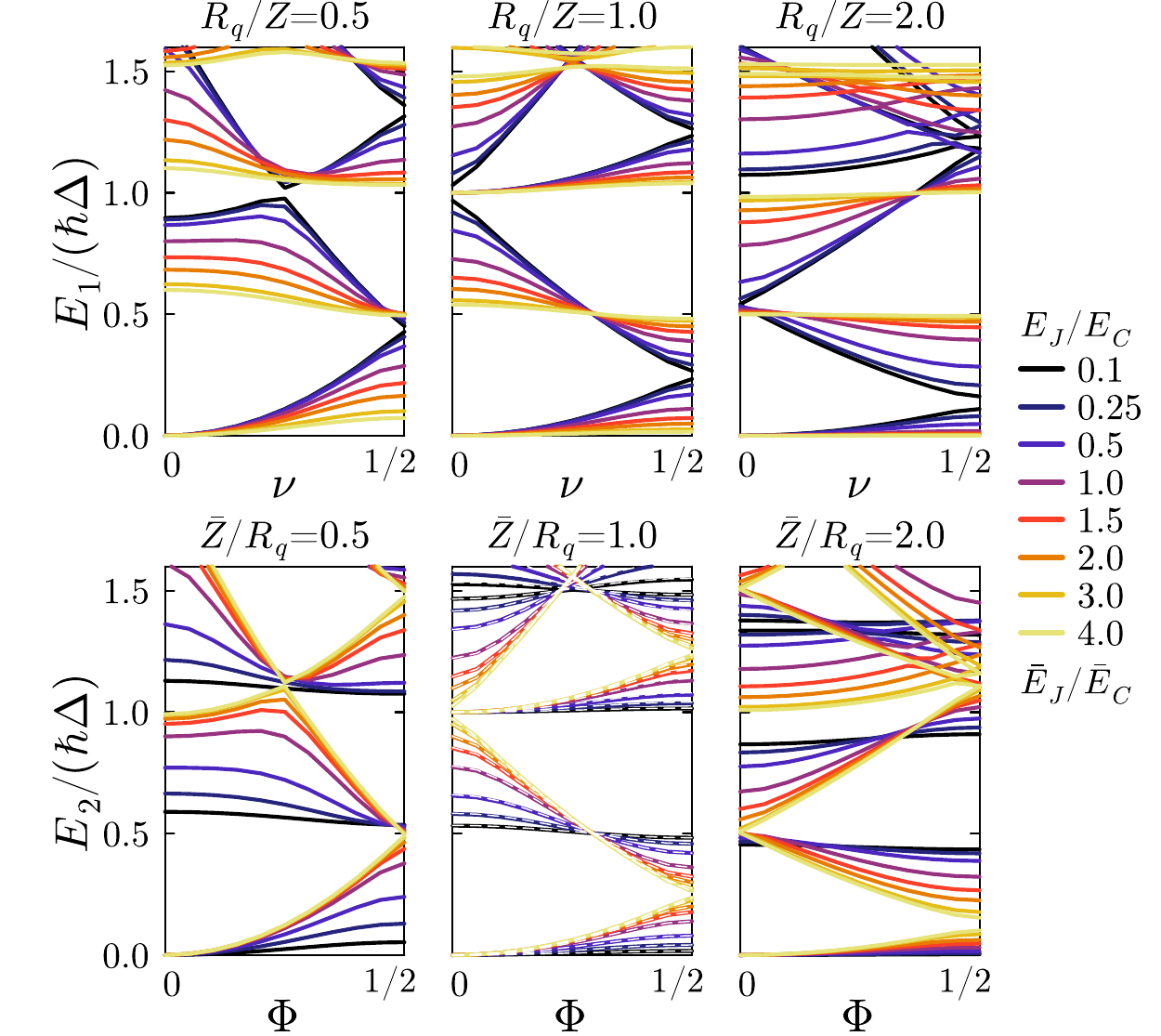}
\caption{Plots analogous to Fig.3(a)-(b) for two additional impedance valies. Upper panels: charge-dependent energy band dispersions of the capacitively terminated circuit, shown for three different impedances and various values of the ratio $E_J/E_C$.  
Lower panels: corresponding flux-dependent energy bands for the inductively terminated circuit.  
Thin dashed lines indicate the predictions from the conformal field theory formula~\eqref{eq:cft-spectrum}, where the parameter $\mu$ has been fitted to the numerical data.  
Throughout, we fix $\hbar \omega_p = 4E_C$ and $\hbar \Delta = 0.66E_C$.
}
\label{fig:EJ-dependence-three}
\end{figure}

In Fig.~\ref{fig:EJ-dependence-three}, we display two the $E_J$ dependence of the spectra of the two circuits shown in Fig.3(a)-(b) for two additional impedance values.

\end{document}